\documentclass[12pt]{article}

\usepackage{a4wide}
\usepackage[dvips]{graphicx}

\def\eqcm{\: ,}           % punctuation in equations
\def\eqpt{\: .}
\def\adj{{\phantom{.}}}   % alignment of subscripts
\def\nn{\nonumber}
\newcommand{\ovl}[1]{\overline{#1}}
\newcommand{\lcvec}[3]{\left[\;#1\;,\;#2\;,\;#3\;\right]}
\renewcommand{\d}{{\rm d}}

\begin{document}

\begin{flushright}

SLAC-PUB-8613\\
WUB 00-10 \\
PITHA 00/21\\
hep-ph/0009255 \\
September 2000
\end{flushright}

\renewcommand{\thefootnote}{\fnsymbol{footnote}}

\begin{center}
\vskip 3.5\baselineskip
\textbf{\Large The Overlap Representation of Skewed \\[0.3\baselineskip]
Quark and Gluon Distributions\footnote{Work supported by Department
of Energy contract DE--AC03--76SF00515.}}
\vskip 2.5\baselineskip
M.~Diehl$^{1}$\footnote{Supported by the Feodor Lynen Program of the
Alexander von Humboldt Foundation.}, 
Th.~Feldmann$^{2}$,
R.~Jakob$^{3}$\footnote{Supported by the Deutsche
Forschungsgemeinschaft.} 
\ and P.~Kroll$^{3}$
\vskip \baselineskip
1. Stanford Linear Accelerator Center, Stanford University, Stanford,
   CA 94309, U.S.A.
2. Institut f\"ur Theoretische Physik E, RWTH Aachen, 52056 Aachen,
   Germany \\  
3. Fachbereich Physik, Universit\"at Wuppertal, D-42097 Wuppertal,
   Germany
%
%\vskip \baselineskip
%{\large September 7, 2000}
%
\vskip 3\baselineskip
\textbf{Abstract}\\[0.5\baselineskip]
\parbox{0.9\textwidth}{Within the framework of light-cone quantisation
we derive the complete and exact overlap representation of skewed
parton distributions for unpolarised and polarised quarks and
gluons. Symmetry properties and phenomenological applications are
discussed.}
\vskip 1.5\baselineskip
\textit{Submitted to Nuclear Physics B}
\vskip 1.5\baselineskip
\end{center}

\renewcommand{\thefootnote}{\arabic{footnote}}
\setcounter{footnote}{0}

% end of title page

%%%%%%%%%%%%%%%%%%%%%%%%%%%%%%%%%%%%%%%%%%%%%%%%%%%%%%%%%%%%%%%%%%%%%%%%
\section{Introduction}
\label{sec:introduction}
%%%%%%%%%%%%%%%%%%%%%%%%%%%%%%%%%%%%%%%%%%%%%%%%%%%%%%%%%%%%%%%%%%%%%%%%

Quark and gluon fields are the basic objects of Quantum Chromodynamics
(QCD), the theory of strong interactions. On the other hand, the
objects observable in experiment are hadrons, which are built up from
quarks and gluons. There is still no analytical description of the
mechanism of confinement which relates hadron properties to the quark
and gluon degrees of freedom. In hard scattering processes, instead,
we parameterise the non-perturbative information in terms of hadronic
matrix elements of quark and gluon field operators. Universality of
the matrix elements assures that once they have been measured in a
hard process, the results can be plugged into the calculation of
observables for other hard processes.

In the description of inclusive and exclusive hard processes the
hadronic matrix elements of quark and gluon fields generally are of
two distinct types:
\begin{itemize}
\item
In inclusive hard processes the hadron state expectation values, i.e.,
\emph{diagonal}\ matrix elements, of \emph{bi-local}\ combinations of
field operators at a light-like distance are conveniently expressed in
terms of parton distribution functions (PDFs)~\cite{Soper:1977jc}. At
leading order in a twist expansion the PDFs acquire a simple intuitive
interpretation in the context of light-cone (LC) quantisation,
provided a light-cone gauge, for instance $A^+=0$, is used. In that
case, the PDFs give the probability to find a parton inside a hadron,
carrying the plus-component fraction $x$ of the hadron's momentum
$p$. Projections on definite helicity or transverse spin eigenstates
of the partons reveal information on their preferences to be aligned
or anti-aligned relative to the spin of the parent hadron. The
polarised quark PDFs, like the helicity distribution $\Delta q(x)$ or
the transverse spin distribution $\delta q(x)$, thus have an
interpretation as differences of probabilities.
\item
The other type of hadronic matrix elements of quark and gluon fields
is involved in the description of hadron form factors at large
momentum transfers. They are defined from matrix elements of
\emph{local}\ combinations of quark field operators. In contrast to the
above mentioned matrix elements of the inclusive processes, the matrix
elements involved in the definition of form factors are
\emph{non-diagonal}\ in initial and final state, which are
characterised by different momenta.
\end{itemize} 

The interest in the connection between hard inclusive and exclusive
reactions has recently been renewed in the context of the so-called
skewed parton distributions
(SPDs)~\cite{Muller:1994fv,Ji:1997ek,Radyushkin:1997ki} which have
been shown to play a decisive role in the understanding of deeply
virtual exclusive reactions, Compton
scattering~\cite{Radyushkin:1997ki,Ji:1998xh} and electroproduction of
mesons~\cite{Radyushkin:1996ru}. The SPDs, defined as 
{\em non-diagonal}\ hadronic matrix elements of {\em bi-local}\ products of
quark and gluon field operators, are functions of the momentum
fraction variable $\bar{x}$, the skewedness parameter $\xi$, and the
squared momentum transfer $t$ (see Sect.~\ref{sec:fock} for
definitions). They represent generalisations of the two types of
hadronic matrix elements mentioned above and are, in so far, hybrid
objects which combine properties of ordinary PDFs and of form
factors. In fact, the close connection of these quantities becomes
manifest in reduction formulas: PDFs are the forward limits of the
SPDs, while form factors are moments of them.

In a recent publication~\cite{Diehl:1999kh} we have obtained a
representation of quark SPDs in the regions $\xi<|\bar x|<1$ in terms
of light-cone wave function (LCWF) overlaps. This representation can
be viewed as a generalisation of the famous Drell-Yan
formula~\cite{Drell:1970km} for electromagnetic form factors. It also
includes the LCWF representation of the PDFs~\cite{Brodsky:1989pv} as
a limiting case. The purpose of the present article is to present the
derivation of the overlap representation for SPDs in detail within the
framework of the light-cone (LC) quantisation. We will generalise our
previous result~\cite{Diehl:1999kh} to the entire kinematical region,
$-1<\bar x<1$, and extend it to the gluonic sector.

The paper is organised as follows: In Sect.~\ref{sec:fock} we recall a
few basic facts of LC quantisation, discuss the Fock state
decomposition of hadronic states, and introduce the relevant
kinematical definitions. The derivation of the overlap representation
for the unpolarised quark SPDs is presented in some detail in
Sect.~\ref{sec:overlap}. The case of polarised quark SPDs and the
extension to the case of gluons is discussed in
Sect.~\ref{sec:polglue}. Sect.~\ref{sec:properties} is devoted to the
discussion of general properties of the overlap representation, and
Sect.~\ref{sec:applications} to phenomenological applications. 
We conclude the article with a summary (Sect.~\ref{sec:summary}). In
an appendix we will present some useful formulas in order to
facilitate the comparison with other conventions in the definition of
SPDs.

%%%%%%%%%%%%%%%%%%%%%%%%%%%%%%%%%%%%%%%%%%%%%%%%%%%%%%%%%%%%%%%%%%%%%%%%
\section{Fock state decomposition}
\label{sec:fock}
%%%%%%%%%%%%%%%%%%%%%%%%%%%%%%%%%%%%%%%%%%%%%%%%%%%%%%%%%%%%%%%%%%%%%%%%

In this Section, after giving some necessary definitions for field
operators and parton states, we discuss the Fock state decomposition
of a hadronic state, a crucial step towards an overlap representation
of SPDs. We will use the component notation 
$z^\mu=\left[z^+,z^-, {\bf z}_\perp\right]$ for any four-vector $z$ 
with the LC components $z^\pm=(z^0\pm z^3)/\sqrt{2}$ and the transverse 
part ${\bf z}_\perp=(z^1,z^2)$. We will work in the framework of LC 
quantisation in the $A^+=0$ gauge. At given light-cone time, say $z^+=0$, the
independent dynamical fields are the so-called ``good'' LC components
of the fields, namely $\phi_q^{\,c}\equiv P_+\psi_q^{\,c}$ for quarks
of flavour $q$ and colour~$c$ (where the projectors acting on the
Dirac field $\psi_q^{\,c}$ are defined by 
$P_\pm=\frac{1}{2}\gamma^\mp\gamma^\pm$) and the transverse components of the
gluon potential $A_\alpha^{\,c}$ (where $\alpha\in\{1,2\}$ is a
transverse index and $c$ again denotes colour).
The independent dynamical fields have Fourier expansions in momentum
space (see e.g.~\cite{Brodsky:1989pv}, Appendix~II)
\begin{eqnarray}
\label{eq:quarkexpansion}
\phi_q^{\,c}(z^-,{\bf z}_\perp)&=&
\int\frac{\d k^+\,\d^2{\bf k}_{\perp}}{k^+\,16\pi^3}\,\Theta(k^+)\;
\sum_{\mu}\;\nn\\ 
&& \bigg\{ \phantom{{} + {}}
     b_q(w)\;u_+(w)
  \;\exp\left({}-i\,k^+z^-+i\,{\bf k}_{\perp}
    \cdot{\bf z}_\perp)\right)
\nn\\
&& \phantom{\bigg\{ }
{{} + {}} d_q^{\,\dagger}(w)\,v_+(w)
  \;\exp\left({}+i\,k^+z^--i\,
    {\bf k}_{\perp}\cdot{\bf z}_\perp)\right)
\bigg\} 
\end{eqnarray} 
for the free quark field, and ($\alpha\in\{1,2\}$)
\begin{eqnarray} 
\label{eq:gluonexpansion}
A_\alpha^{\,c}(z^-,{\bf z}_\perp) &=& 
\int\frac{\d k^+\,\d^2{\bf k}_{\perp}}{k^+\,16\pi^3}\,\Theta(k^+)\;
\sum_{\mu}  \nn\\
&& \bigg\{ \phantom{{} + {}}
     a^\adj(w)\;\epsilon_\alpha(w)\;
     \exp\left({}-i\,k^+z^-+i\,
         {\bf k}_{\perp}\cdot{\bf z}_\perp\right)
\nn\\
&& \phantom{\bigg\{ }
{{} + {}} a^{\,\dagger}(w)\,\epsilon^*_\alpha(w)\;
     \exp\left({}+i\,k^+z^--i\,
         {\bf k}_{\perp}\cdot{\bf z}_\perp\right)
\bigg\} 
\eqpt 
\end{eqnarray} 
for the free gluon field, where $\Theta(k^+)$ denotes the usual step
function. We use a collective notation for the dependence on the plus
and transverse parton momentum components, the helicity, and the
colour in the form
\begin{equation} 
w=(k^+,{\bf k}_{\perp}, \mu, c)
\eqpt
\label{eq:w-collective}
\end{equation} 
The operators $b$ and $d^\dagger$ are the annihilator of the ``good''
component of the quark fields and the creator of the ``good''
component of the antifields, respectively, and $u_+(w)\equiv
P_+\,u(w)$ and $v_+(w)\equiv P_+\,v(w)$ are the projections of the
usual quark and antiquark spinors. $a$ and $a^\dagger$ are the
annihilation and creation operators for transverse gluons, and
$\epsilon_\alpha(w)$ is a transverse component of the gluon
polarisation vector. The operators fulfil the anticommutation
relations
\begin{eqnarray} 
\label{eq:anticommutation}
\left\{b_{q'}^\adj(w'),b_{q}^{\,\dagger}(w)\right\}&=&
\left\{d_{q'}^\adj(w'),d_{q}^{\,\dagger}(w)\right\}\nn\\
&=&
16\pi^3\;k^+\;\delta(k^{\prime\,+}-k^{+})\;
\delta^{(2)}\left({\bf k}'_{\perp}-{\bf k}_{\perp}\right)\;
\delta_{q'q} \;\delta_{\mu'\mu} \;\delta_{c'c} 
\eqcm
\end{eqnarray} 
and the gluon operators $a$ and $a^\dagger$ satisfy the commutation
relation
\begin{equation} 
\label{eq:commutation}
\left[a^\adj(w'),a^{\,\dagger}(w)\right]=
16\pi^3\;k^+\;\delta(k^{\prime\,+}-k^{+})\;
\delta^{(2)}\left({\bf k}'_{\perp}-{\bf k}_{\perp}\right)\;
\delta_{\mu'\mu}\;\delta_{c'c}
\eqpt
\end{equation} 

Key ingredient in our derivation of the overlap representation is the
Fock state decomposition~\cite{Brodsky:1989pv}, i.e., the replacement
of a hadron state by a superposition of partonic Fock states
containing free quanta of the ``good'' LC components of (anti)quark
and gluon fields.  Single-parton, quark, antiquark or gluon, momentum
eigenstates are created by $b^\dagger$, $d^\dagger$ and $a^\dagger$
acting on the perturbative vacuum,\footnote{We assume a `trivial'
perturbative vacuum, i.e., $b\,|0\rangle=d\,|0\rangle=a\,|0\rangle=0$,
and ignore possible problems arising from zero modes, which are beyond
the scope of this investigation.}
\begin{eqnarray}
\label{eq:partonstate} 
|q;w \rangle &=& 
b_q^\dagger (w) \, |0\rangle \eqcm
\nn\\
|\bar q; w \rangle &=&
d_q^\dagger (w) \, |0\rangle \eqcm
\nn\\
|g; w \rangle &=&
a^\dagger (w) \, |0\rangle  \eqcm
\end{eqnarray}  
and the (anti)commutation relations (\ref{eq:anticommutation}) and 
(\ref{eq:commutation}) translate into the normalisation of these states,
\begin{equation}
\label{eq:statenorm}
\langle s'; w'|s;w\rangle=
16\pi^3\,k^+\,\delta(k^{\prime\,+}-k^+)\, 
\delta^{(2)}\left({\bf k}'_{\perp}-{\bf k}^\adj_{\perp}\right)
\delta_{s's}\;\;\delta_{\mu'\mu}\;\delta_{c'c}
\end{equation}
for partons $s$, $s'$ of any kind. A hadronic state characterised by
the momentum $p$ and helicity $\lambda$ is written as
\begin{equation}
\label{eq:Fockstate}
\left|H;p,\lambda\right\rangle = \sum_{N,\beta} 
\int [\d x]_N [\d^2 {\bf k}_\perp]_N\;
\Psi_{N,\beta}^\lambda(r) \;
\left|N,\beta;k_1,\ldots,k_N \right\rangle \eqcm
\end{equation} 
where $\Psi_{N,\beta}^\lambda(r)$ is the momentum LCWF of the
$N$-parton Fock state $|N,\beta;k_1,\ldots,k_N \rangle$. The index
$\beta$ labels its parton composition, and the helicity 
and colour of each parton.

Apart from their discrete quantum numbers (flavour, helicity, colour)
the partons are characterised by their momenta
$k_i=\lcvec{k_i^+}{k_i^-}{{\bf k}_{\perp i}}$. The LCWFs, on the other
hand, do not depend on the momentum of the hadron, but only on the
momentum coordinates of the partons \emph{relative}\ to the hadron
momentum. In other words, the centre of mass motion can be separated
from the relative motion of the partons~\cite{Dirac:1949cp}. The
arguments of the LCWF, namely $x_i\equiv k_i^+/p^+$ and the transverse
momenta ${\bf k}_{\perp i}$, can most easily be identified in
reference frames where the hadron has zero transverse momentum. We
call such frames ``hadron-frames'' and use again a collective notation
\begin{equation}
\label{eq:r-collective}
r_i=(x_i,{\bf k}_{\perp i})
\end{equation} 
and $\Psi_{N,\beta}^\lambda(r)=\Psi_{N,\beta}^\lambda(r_1,\ldots,r_N)$
for the arguments of the LCWFs.\footnote{This notation resembles the
definition of the $w$ in (\ref{eq:w-collective}), but refers now to
the {\em relative}\ momentum coordinates.}  An $N$-parton state is
defined as
\begin{equation}
\label{parton-states}
\left|N,\beta;k_1,\ldots,k_N \right\rangle = 
\frac{1}{\sqrt{f_{N,\beta}}}\,
\prod_i \frac{b_{q_i}^{\,\dagger}(w_i)}{\sqrt{x_i}}\;\;
\prod_j \frac{d_{q_j}^{\,\dagger}(w_j)}{\sqrt{x_j}}\;\;
\prod_l \frac{a^{\,\dagger}(w_l)}{\sqrt{x_l}} \;|0\rangle
\eqpt
\end{equation} 
Owing to the (anti)commutation relations (4) and (5) the states
$|N,\beta;k_1,\ldots,k_N\rangle$ are completely (anti)symmetric under
exchange of the momenta $k_i$ of gluons (quarks) with identical
discrete quantum numbers.\footnote{Notice that this is different from
the notation of quantum mechanics, used e.g.\ in \cite{Diehl:1999kh},
where $|s_1,w_1;s_2,w_2\rangle$ is defined as a direct product
$|s_1,w_1\rangle\otimes |s_2,w_2\rangle$ and therefore different from
$\pm |s_2,w_2; s_1,w_1\rangle$.} Without loss of generality we can
thus take the wave functions $\Psi_{N,\beta}^\lambda(r)$ to have the
same (anti)symmetry under permutations of the corresponding momenta
$r_i$. The normalisation constant $f_{N,\beta}$ in
(\ref{parton-states}) contains a factor $n!$ for each subset of $n$
partons whose discrete quantum numbers are identical, so that one has
\begin{eqnarray}
\label{eq:parton-state-norm}
\lefteqn{
\Psi_{N,\beta'}^{*\,\lambda}(r')\, \Psi_{N,\beta}^\lambda(r) \;
\left\langle N',\beta';k'_1,\ldots,k'_{N'} \right.
       \left|N,\beta;k_1^\adj,\ldots,k^\adj_N \right\rangle
}
\nn\\
&=& |\Psi_{N,\beta}^\lambda(r)|^2\;
\delta_{N' N}\, \delta_{\beta'\beta}\;
    \prod_{i=1}^N 16\pi^3\,k^+_i\,\delta(k^{\prime +}_i-k^{+}_i)\, 
\delta^{(2)}\left({\bf k}'_{\perp i}-{\bf k}^\adj_{\perp i}\right)
\eqpt
\end{eqnarray}
The Kronecker symbol $\delta_{\beta'\beta}$ implies that one does not
introduce different labels $\beta$ for states whose assignment of
discrete quantum numbers for the individual partons only differs by a
permutation. Finally, the hadron states are normalised as
\begin{equation}
\label{eq:hadronnorm}
\left\langle H; p',\lambda'\right.\left|H;p,\lambda\right\rangle=
16\pi^3\,p^+\,\delta(p^{\prime +}-p^{+})\, 
\delta^{(2)}\left({\bf p}'_{\perp}-{\bf p}^\adj_{\perp}\right)
\;\delta_{\lambda'\lambda}
\eqcm
\end{equation}
with 
\begin{equation}
\label{eq:Psinorm}
\sum_{N,\beta} \int [\d x]_N [\d^2 {\bf k}_\perp]_N\;
|\Psi_{N,\beta}^\lambda(r)|^2=1
\eqpt
\end{equation} 
The integration measures in Eqs.~(\ref{eq:Fockstate}) and
(\ref{eq:Psinorm}) are defined by
\begin{eqnarray}
[{\rm d}x]_N &\equiv& \prod_{i=1}^N {\rm d}x_i\;
     \delta\!\left(1-\sum_{i=1}^N x_i\right) \eqcm \\ 
{} [{\rm d}^2{\bf k}_\perp]_N
&\equiv& 
\frac{1}{(16\pi^3)^{N-1}}\,\prod_{i=1}^N 
     {\rm d}^2 {\bf k}_\perp{}_i \; 
     \delta^{(2)}\!\left(\sum_{i=1}^N {\bf k}_\perp{}_i-\bf{p}_\perp\right) 
\eqpt
\end{eqnarray}
We remark that the parton states (\ref{parton-states}) do not refer to
a specific hadron, rather they are characterised by a set $\beta$ of
flavour, helicity and colour quantum numbers. Their coupling to a
colour singlet hadron with definite quantum numbers such as isospin is
incorporated in the LCWFs $\Psi_{N,\beta}^\lambda(r)$. Many of them
are zero, and several of the non-zero ones are related to each
other. The three-quark (valence) Fock state of the nucleon, for
instance, has only one independent LCWF for all configurations where
the quark helicities add up to the helicity of the nucleon
\cite{Dziembowski:1988es}. For higher Fock states there are in general
several independent LCWFs.

An alternative and frequently used method is to use parton states that
are colour neutral and already have the appropriate quantum numbers of
the hadron. These parton states are linear combinations of our states
(\ref{parton-states}). The 
LCWFs in this method correspond to the independent ones of our
coupling scheme, up to a normalisation factor. The label $\beta$ is
restricted correspondingly. As we will see below, our method has
advantages in the central region, $-\xi<\bar x<\xi$, while it is
equivalent in the other regions of $\bar x$.

Since hadrons are not massless, we need to specify the helicity states
appearing in the Fock state decomposition (\ref{eq:Fockstate}). To
this end we briefly review its construction. One introduces the wave
functions $\Psi_{N,\beta}^\lambda(r)$ by first writing down
(\ref{eq:Fockstate}) for a state with ${\bf p}_{\perp} = 0$, i.e., in
a ``hadron frame''. There, helicity states $|H;p,\lambda\rangle$ are
defined in the usual way, the spin direction being aligned or
antialigned with the hadron momentum. One then obtains the Fock state
decomposition for a hadron with nonzero ${\bf p}_{\perp}$ by applying
to the states on both sides of (\ref{eq:Fockstate}) a ``transverse
boost'' (see e.g.~\cite{Brodsky:1989pv}), i.e., a transformation that
leaves the plus component of {\em any}\ four-vector $z$ unchanged. It
involves the parameters $b^+$ and ${\bf b}_\perp$ and reads
\begin{equation} 
\label{eq:plustrafo} 
\lcvec{z^+}{z^-}{{\bf z}_\perp}
\qquad\longrightarrow\qquad
\lcvec{z^{+}}{z^{-}-\frac{{\bf z}_\perp\cdot{\bf b}_\perp}{b^+}
+\frac{z^+\,{\bf b}_\perp^{\,2}}{2\,(b^+)^2}}
      {{\bf z}_\perp-\frac{z^+}{b^+}\,{\bf b}_\perp} \eqpt
\end{equation}
This transformation relates the parton momenta
in the frame where ${\bf p}_{\perp} \neq 0$ with those, $r_i$, in the
hadron frame. It also specifies the spin state of the hadron. One
easily sees that its covariant spin vector reads
\begin{equation}
\frac{\lambda}{m}\, 
     \lcvec{p^+}{\frac{{\bf p}_{\perp}^2 - m^2}{2p^+}}{{\bf p}_{\perp}} 
\end{equation}
and, as remarked in \cite{Diehl:1999kh}, is a linear combination of
the four-vectors $p$ and $\lcvec{0}{1}{{\bf 0}_{\perp}}$. In the limit
$m=0$, the light-cone helicity states so defined coincide with the
usual helicity states. With finite $m$ they do not: for ${\bf
p}_{\perp} \neq 0$ they are not eigenstates of the angular momentum
operator along their direction of motion, unless one goes to the
infinite-momentum frame \cite{Kogut:1970xa}. Explicit spinor
representations can e.g.\ be found
in~\cite{Brodsky:1989pv}.\footnote{The spinors given by Brodsky and
Lepage~\cite{Brodsky:1989pv} are equivalent to those proposed by Kogut
and Soper \cite{Kogut:1970xa} if one takes into account the difference
in the conventions for light-cone coordinates and in the
representations of the Dirac matrices.}

\begin{figure}[tb] 
\begin{center} 
\includegraphics[width=64mm]{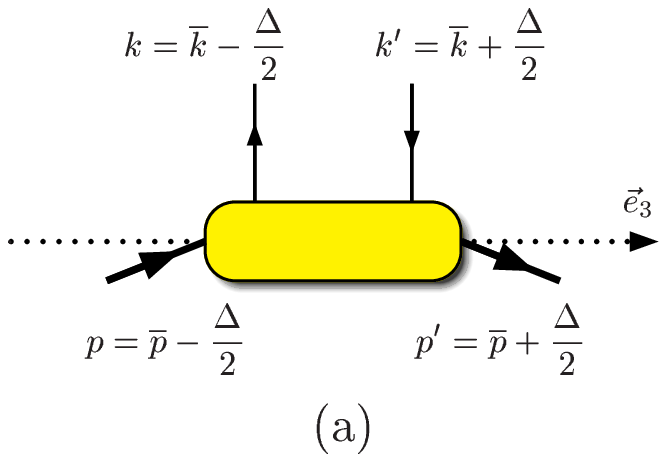}
\qquad
\includegraphics[width=64mm]{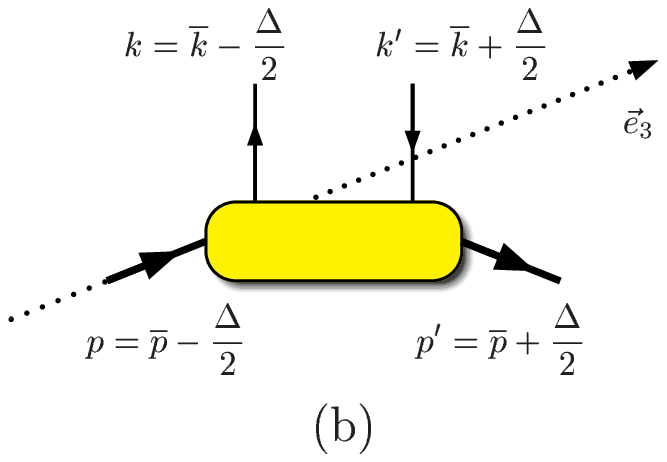}
\end{center} 
\caption{Illustration of two common choices to fix the longitudinal
direction in the definition of the non-diagonal hadronic matrix
element which defines skewed parton distributions. The flow of momenta
is indicated on the lines.}
\label{fig:one}
\end{figure}

Let us now take a look at the hadron momenta involved in the
definitions of SPDs. The initial and final hadron states are
characterised by the momenta $p$ and $p^{\,\prime}$. In order to
parameterise them we define the average momentum
\begin{equation}
\bar p= \frac{1}{2}\, (p+p^{\,\prime}) \eqcm
\label{eq:averagemom}
\end{equation}
choose the three-momentum ${{\bf\bar p}}$ to 
be along the ${\bf e}_{\,3}$-axis (see Fig.~\ref{fig:one}), and write in
light-cone components 
\begin{eqnarray} 
\label{eq:ji-parameterisation}
p  &=& 
\lcvec{(1+\xi)\bar p^{\,+}}
      {\frac{m^2+{\bf\Delta}_\perp^2/4}{2(1+\xi)\,\bar p^{\,+}}}
      {-\frac{{\bf\Delta}_\perp}{2}} \eqcm \nn \\
p^{\,\prime} &=& 
\lcvec{(1-\xi)\,\bar p^{\,+}}
      {\frac{m^2+{\bf\Delta}_\perp^2/4}{2(1-\xi)\,\bar p^{\,+}}}
      {+\frac{{\bf\Delta}_\perp}{2}}
\end{eqnarray}
with the transverse vector 
${\bf\Delta}_\perp$, the plus momentum $\bar p^{\,+}$,
and the skewedness parameter
\begin{equation}
\label{eq:skewedness}
\xi= \frac{(p-p^{\,\prime})^+}{(p+p^{\,\prime})^+}
\eqcm
\end{equation}
which describes the change in plus momentum. The momentum transfer
takes the form
\begin{equation} 
\label{eq:Delta}
\Delta=p^{\,\prime}-p=
\lcvec{-2\xi\,\bar p^{\,+}}
      {\frac{\xi (m^2+{\bf\Delta}_\perp^2/4)}{(1-\xi^2)\bar p^{\,+}}}
      {{\bf\Delta}_\perp} \eqcm
\end{equation} 
and with the parameterisation (\ref{eq:ji-parameterisation}) its
square reads
\begin{equation} 
t=\Delta^2= - \frac{4\,\xi^2\,m^2 + {\bf\Delta}_\perp^2}{1-\xi^2} \eqpt
\end{equation} 
Notice that the positivity of ${\bf\Delta}_\perp^2$ implies a minimal value
\begin{equation}
  -t_0= \frac{4\xi^2 m^2}{1-\xi^2}
\label{eq:t-nought}
\end{equation}
for $-t$ at given $\xi$, which translates into a maximum allowed $\xi$
at given $t$. 
As shown in Fig.~\ref{fig:one}(a) the parton emitted by the hadron has the
momentum $k$, and the one absorbed has momentum $k'$. The average
parton momentum $\bar k$ is defined as $(k+k')/2$, in analogy to
(\ref{eq:averagemom}), and correspondingly a momentum fraction $\bar
x=\bar k^+/\bar p^+$ is introduced. This is Ji's variable
$x$~\cite{Ji:1997ek}.

In an alternative parametrisation of the hadron momenta frequently
found in the literature (see for instance~\cite{Radyushkin:1997ki}),
the three-momentum of the incoming proton is chosen to lie along the
${\bf e}_{\,3}$-axis. In Fig.~\ref{fig:one} the two different choices
are illustrated. We will present our results formulated in terms of
the alternative set of kinematical variables in the Appendix.

%%%%%%%%%%%%%%%%%%%%%%%%%%%%%%%%%%%%%%%%%%%%%%%%%%%%%%%%%%%%%%%%%%%%%%%%%
\section{The unpolarised skewed quark distribution}
\label{sec:overlap}
%%%%%%%%%%%%%%%%%%%%%%%%%%%%%%%%%%%%%%%%%%%%%%%%%%%%%%%%%%%%%%%%%%%%%%%%%
We now turn to the derivation of the overlap representation for
leading-twist SPDs. For definiteness let us consider the case of
unpolarised quarks inside protons. Thus, we investigate the proton
matrix element of the plus component of a flavour-diagonal bi-local
quark field operator summed over colour. The generalisation to other
hadrons is straightforward. Following Ji~\cite{Ji:1997ek}, we define
the SPDs $H^q(\bar x,\xi;t)$ and $E^q(\bar x,\xi;t)$ for a quark of
flavour $q$ by
\begin{eqnarray} 
\label{eq:quarkSPDdef}
{\cal H}_{\lambda'\lambda}^q &\equiv&
\frac{1}{2\sqrt{1-\xi^2}} \;
\sum_c
\int\frac{\d z^-}{2\pi}\;e^{i\,\bar x\,\bar p^{\,+}z^-}\;
\langle p^{\,\prime},\lambda'|
   \,\ovl\psi_q^{\,c}(-\bar z/2)\,\gamma^+\,\psi_q^{\,c}(\bar z/2)\,
                             |p,\lambda\rangle
\nn\\[1\baselineskip]
&=&
 \frac{\ovl u(p^{\,\prime},\lambda')\gamma^+ u(p,\lambda)}
      {2\bar p^{\,+}\sqrt{1-\xi^2}}\; 
H^q(\bar x,\xi;t)
+\frac{\ovl u(p^{\,\prime},\lambda')
         {i\sigma^{+\alpha}\Delta_\alpha} u(p,\lambda)}
      {4m\,\bar p^{\,+}\sqrt{1-\xi^2}}\; 
E^q(\bar x,\xi;t)
\end{eqnarray} 
where $\lambda$, $\lambda'$ denote the proton helicities, and $\bar z$
is a shorthand notation for $[0,z^-,{\bf 0_\perp}]$. The link operator
normally needed to render the definition (\ref{eq:quarkSPDdef}) 
gauge-invariant does not
appear because we choose the gauge $A^+=0$, which together with an
integration path along the minus direction reduces the link operator
to unity.  With the phase conventions of the Brodsky-Lepage light-cone
spinors~\cite{Brodsky:1989pv} we find for the different proton
helicity combinations
\begin{eqnarray}
\label{eq:quark-helcomb}
{\cal H}^q_{++} = \phantom{-( } {\cal H}^q_{--} \phantom{ )^*} &=& 
      H^q - \frac{\xi^2}{1-\xi^2}\, E^q \eqcm
\nn\\
{\cal H}^q_{-+} = - ( {\cal H}^q_{+-} )^* &=& 
      \eta\, \frac{\sqrt{t_0-t}}{2m}\, \frac{1}{\sqrt{1-\xi^2}}\,
      E^q 
\end{eqnarray}
with $t_0$ defined in Eq.~(\ref{eq:t-nought}) and a phase factor reading
\begin{equation}
\label{eq:phase}
  \eta = \frac{\Delta^1 + i \Delta^2}{|{\bf \Delta}_\perp|}
\end{equation}
for proton momenta of the form (\ref{eq:ji-parameterisation}). In a
general reference frame $\Delta^\alpha$ in Eq.~(\ref{eq:phase}) is to
be replaced with $\Delta^\alpha - (\Delta^+/\bar{p}^{\,+})\,
\bar{p}^{\,\alpha}$. Evaluating ${\cal H}_{\lambda'\lambda}^q$ for
both proton helicity flip and non-flip, one obtains the usual SPDs for
quark flavour $q$, $H^q$ and $E^q$. We remark in passing that the
LCWFs for proton helicity states with opposite helicity are related
through a reflection about the $x$-$z$ plane
\cite{Dziembowski:1988es}.

\begin{figure}[hbt] 
\begin{center} 
\includegraphics[width=0.9\textwidth]{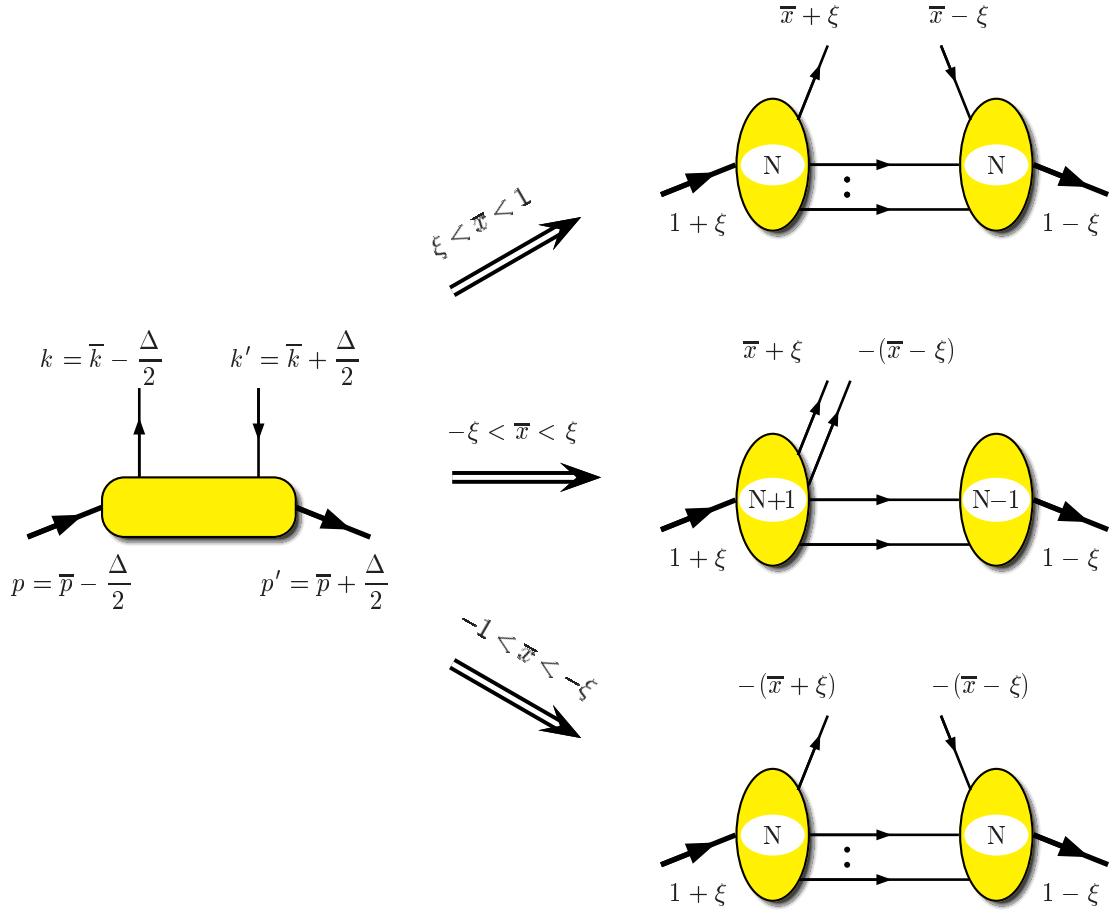}
\end{center} 
\caption{Overlap representations for SPDs in different kinematic
regions for the case $\xi>0$. The flow of momenta is indicated on the
lines.  Top (bottom) right: the region $\xi<\bar x<1$ ($-1<\bar
x<-\xi$), where the SPDs are given by $N\to N$ overlaps.  Middle
right: the central region $-\xi<\bar x<\xi$, where $N+1\to N-1$
overlaps are relevant.}
\label{fig:two}
\end{figure}

A key point in our derivation of an overlap formula is the well-known
observation that the bi-local quark field operator in the definition
(\ref{eq:quarkSPDdef}) can be written as a density operator in terms
of the ``good'' LC components (see e.g.~\cite{Jaffe:zw})
\begin{equation} 
\ovl\psi{}_q^{\,c}(-\bar z/2)\,\gamma^+\,\psi_q^{\,c}(\bar z/2)=
\sqrt{2}\;\phi_q^{\,c\,\dagger}(-\bar z/2)\,\phi_q^{\,c}(\bar z/2) .
\end{equation}  
Inserting the momentum space expansion (\ref{eq:quarkexpansion}) of
the fields, one obtains for the Fourier transform occurring in
(\ref{eq:quarkSPDdef})
\begin{eqnarray}
\label{eq:density-expansion}
\lefteqn{
\sum_c
\int\frac{\d z^-}{2\pi}\;e^{i\,\bar x\,\bar p^{\,+}z^-}\;
\ovl\psi{}_q^{\,c}(-\bar z/2)\,
\gamma^+\,\psi_q^{\,c}(\bar z/2) } \hspace{4em}
\nn\\
&=& 
2 \,
\int\frac{\d k'^+\,
          \d^2{\bf k}'_{\perp}}{k'^+\,16\pi^3}\,\Theta(k'^+)\;
\int\frac{\d k^+\,\d^2{\bf k}_{\perp}}{k^+\,16\pi^3}\,\Theta(k^+)\;
\sum_{\mu, \mu',c,c'}\;\delta_{c'c}
\nn\\
&& \Big\{ \:\:\;
       \delta(2\bar{x} \bar{p}^+ - k'^+ - k^+)\;
       b_q^\dagger(w')\, b_q^{\phantom{\dagger}}(w)\, 
       u_+^\dagger(w')\, u_+(w) \nn\\
&& {}+ \delta(2\bar{x} \bar{p}^+ + k'^+ + k^+)\;
       d_q^{\phantom{\dagger}}(w')\, d_q^\dagger(w)\, 
       v_+^\dagger(w')\, v_+(w) \phantom{\Big\}} \nn\\
&& {}+ \delta(2\bar{x} \bar{p}^+ + k'^+ - k^+)\;
       d_q^{\phantom{\dagger}}(w')\, b_q^{\phantom{\dagger}}(w)\, 
       v_+^\dagger(w')\, u_+(w) \phantom{\Big\}} \nn\\
&& {}+ \delta(2\bar{x} \bar{p}^+ - k'^+ + k^+)\;
       b_q^\dagger(w')\, d_q^\dagger(w)\, 
       u_+^\dagger(w')\, v_+(w) \; \Big\} \eqcm
\end{eqnarray}
a form that readily allows one to interpret the SPDs in the parton
picture \cite{Ji:1998pc}. Which of the four terms in
(\ref{eq:density-expansion}) contributes to the matrix element in
(\ref{eq:quarkSPDdef}) is determined by the positivity conditions $k^+
\ge 0$ and $k'^+ \ge 0$ for the parton momenta, together with momentum
conservation, which imposes $k^+ - k'^+ = p^+ - p^{\prime\,+} =
2\xi\bar{p}^{\,+}$. For definiteness we consider the case $\xi>0$ in
the following, which is relevant for the applications of the SPDs in
hard processes that have so far been considered in the literature. In
the region $\xi<\bar x<1$ the SPDs describe the emission of a quark from
the nucleon with momentum fraction $\bar x+\xi$ and its reabsorption
with $\bar x-\xi$. In the region $-1<\bar x<-\xi$ one has the emission
of an antiquark from the nucleon with momentum fraction $-(\bar
x+\xi)$ and its reabsorption with $-(\bar x-\xi)$. In the third region
$-\xi<\bar x<\xi$, however, the nucleon emits a quark-antiquark
pair. We will discuss the three cases separately; first we focus on
the region $\xi<\bar x<1$ (see Fig.\ref{fig:two}). The last term in
(\ref{eq:density-expansion}), going with $b^\dagger(w')\,
d^\dagger(w)$ and describing the absorption of a quark-antiquark pair,
does not contribute for $\xi>0$.

We remark in passing that one can define distributions
$H^{\bar{q}}(x,\xi;t) \equiv -H^q(-x,\xi;t)$ and $E^{\bar{q}}(x,\xi;t)
\equiv -E^q(-x,\xi;t)$, which in the region $\xi<\bar x<1$ describe
the emission and reabsorption of antiquarks and may thus be called
``skewed antiquark distributions''. We will not need this here, and
instead work with the distributions $H^q$ and $E^q$ and their
different interpretation in the three $\bar x$ intervals just
discussed.

%%%%%%%%%%%%%%%%%%%%%%%%%%%%%%%%%%%%%%%
\subsection{The region $\xi<\bar{x}<1$}

The Fock state decomposition (\ref{eq:Fockstate}) leads to a
representation of the matrix element ${\cal H}_{\lambda'\lambda}^q$ as
a sum over contributions from separate Fock states,
\begin{equation}
\label{eq:Focksummation}
{\cal H}_{\lambda'\lambda}^q=
\sum_N {\cal H}_{\lambda'\lambda}^{q(N\to N)} \eqcm
\end{equation}  
with 
\begin{eqnarray}
\label{eq:Hpartonic}
{\cal H}_{\lambda'\lambda}^{q(N\to N)} &=&
\frac{1}{\sqrt{2(1-\xi^2)}}\; 
\sum_c
\sum_{\beta,\beta'} \;
\int [\d \tilde x]_N [\d^2 \tilde{\bf k}_\perp]_N \;
     [\d \hat x']_N [\d^2 \hat{\bf k}'_\perp]_N \;
\Psi_{N,\beta'}^{*\,\lambda'}(\hat r')\,
      \Psi_{N,\beta}^\lambda(\tilde r)
\nn\\
&\times&
\int\frac{\d z^-}{2\pi}\;e^{i\,\bar x\,\bar p^{\,+}z^-} \,
\left\langle N,\beta';k'_1\ldots k'_N\right|
        \,\phi^{\,c\,\dagger}_q(-\bar z/2)\,\phi^{\,c}_q(\bar z/2)\,
        \left|N,\beta;k_1,\ldots,k_N\right\rangle \;
\eqpt \hspace{2em}
\end{eqnarray}
One can now express the $N$-parton states and the bi-local quark field 
operator $\phi^{\,c\,\dagger}_q\,\phi^{\,c}_q$
in terms of the creation and annihilation operators, see 
Eqs.~(\ref{parton-states}) and (\ref{eq:density-expansion}), and evaluate 
the resulting vacuum matrix element using the (anti)commutation relations 
(\ref{eq:anticommutation}) and (\ref{eq:commutation}). 

One obtains a product of two anticommutators involving the creation and
annihilation operators from 
$\phi^{\,c\,\dagger}_q\,\phi^{\,c}_q$, which can 
be rewritten as a matrix element of the field operators for the active 
quarks, and a product of $N-1$ (anti)commutators for the spectator 
partons, which is conveniently expressed through one-parton matrix elements 
as in (\ref{eq:statenorm}). The quantum numbers for the active quarks 
and for the spectators have to match, see Eq.~(\ref{eq:statenorm}), and 
Eqs.~(\ref{eq:activematrix}) and (\ref{spinor-product}) below, so that 
the Fock state labels $\beta$ and $\beta'$ are constrained to be
the same.

For identical partons in the Fock state the non-zero (anti)commutators
generate a number of terms corresponding to the different possibilities to
associate the partons in the initial and final states. 
These terms are, however, all the same because of the (anti)symmetry of
the wave functions  under permutations of the momenta $r_i$ for identical 
particles. The number of these terms equals the
product $\sqrt{f_{N,\beta'}f_{N,\beta}}$ of normalisation factors from the 
parton states times the multiplicity of the active parton. Thus, we end 
up with only one term, a situation which 
allows us to number the spectators in one specific way. We thus arrive at the 
following replacement of the matrix
element appearing in (\ref{eq:Hpartonic})
\begin{eqnarray}
\lefteqn{
\left\langle N,\beta';k'_1\ldots k'_N\right|
        \,\phi^{\,c\,\dagger}_q(-\bar z/2)\,\phi^{\,c}_q(\bar z/2)\,
        \left|N,\beta;k_1,\ldots,k_N\right\rangle}
\nn\\[\baselineskip]
&=&
\sum_{j=1}^N\;
\frac{\langle s'_j;w'_j|
        \;\phi^{\,c\,\dagger}_q(-\bar z/2)\,\phi^{\,c}_q(\bar z/2)\,
                    |s^\adj_j;w^\adj_j\rangle}
{\sqrt{\hat x'_1\ldots \hat x'_N}
      \sqrt{\tilde x_1\ldots \tilde x_N\phantom{\hat x'_N}
      \hspace{-2.5ex}}}\; 
\prod_{i=1\atop i\ne j}^N 
\langle s'_i;w'_i|s^\adj_i;w^\adj_i\rangle\;
\eqpt
\end{eqnarray} 
We now have to comment on the involved parton momenta. The
parton states are characterised by their momenta and helicities,
$w_i$. We denote momenta of partons belonging to the incoming hadron
with unprimed, and the momenta of partons belonging to the outgoing
hadron with primed variables. The LCWFs, on the other hand, depend on
the relative momentum coordinates with respect to the parent hadron,
$r_i$. As mentioned above, the identification of the arguments of the
LCWFs is most easily done when hadron frames are chosen as frames of
reference. We introduce the names ``hadron-in'' and ``hadron-out'' for
frames where the incoming and outgoing hadron has zero transverse
momentum, respectively. For the sake of clarity, we will here and in
the following pedantically label quantities in the hadron-in
(hadron-out) frame with an additional tilde (hat). We further use the
name ``average-frame'' for a system where the hadron momenta are
parameterised in the form of Eqs.~(\ref{eq:ji-parameterisation}).  In
order to achieve a formulation symmetric in incoming and outgoing
quantities it is useful to define as auxiliary variables the averages
of incoming and outgoing parton momenta in the average frame
\begin{equation}
\bar k^\adj_i=\frac{1}{2}\, (k^\adj_i+k'_i) \;,\qquad
\bar x^\adj_i=\frac{\bar k_i^+}{\bar p^{+}}
\eqcm
\label{eq:auxiliary}
\end{equation}
which satisfy
\begin{equation}
\sum_{i=1}^N \bar x_i= \frac{1}{\bar{p}^+}\, 
\sum_{i=1}^N \bar k^{\,+}_i =1
\;,\qquad
\sum_{i=1}^N \bar{\bf k}_{\perp i}=\bar{\bf p}_\perp
= {\bf 0}_\perp
\eqpt 
\end{equation} 
The parton emitted and later reabsorbed from the hadron is called the
``active'' parton and labelled with index $j$; all other partons
$i\neq j$ play the role of ``spectators''.  The active parton carries
a fraction $\bar x_j+\xi$ of the average plus-momentum $\bar{p}^{\,+}$
when it is taken out of the proton, and a fraction $\bar x^\adj_j-\xi$
when it is reinserted. The transverse momentum of the active parton is
${\bf k}_{\perp j}=\bar{\bf k}_{\perp j}-{\bf\Delta}_\perp/2$ before,
and ${\bf k}'_{\perp j}=\bar{\bf k}^\adj_{\perp
j}+{\bf\Delta}^\adj_\perp/2$ after the partonic scattering process.

The arguments of the LCWF for the incoming hadron are obtained through
a transverse boost (\ref{eq:plustrafo}) with parameters
$b^+=(1+\xi)\,\bar p^{\,+}$ and ${\bf b}_\perp=-{\bf\Delta}_\perp/2$,
which leads from the average-frame to the hadron-in frame. Likewise, a
transverse boost with parameters $b^+=(1-\xi)\,\bar p^{\,+}$ and ${\bf
b}_\perp=+{\bf\Delta}_\perp/2$ leads from the average-frame to the
hadron-out frame. {}From momentum conservation and the spectator
condition
\begin{equation} 
k'_i=\bar k^\adj_i=k^\adj_i \eqcm
\qquad \mbox{for }i\neq j 
\eqcm
\end{equation} 
one obtains that the LCWF arguments for the incoming hadron (i.e., the
momenta of the partons belonging to the incoming hadron in the
hadron-in frame) are related to the momenta in the average-frame by
\begin{eqnarray}
\label{eq:tilde-args}
\tilde x_i= \frac{\bar x_i}{1+\xi} \eqcm &\qquad& 
\tilde{\bf k}_{\perp i}=\bar{\bf k}_{\perp i}
                        +\frac{\bar x_i}{1+\xi}\,
                         \frac{{\bf\Delta}_\perp}{2} \eqcm
\qquad \qquad \mbox{for }i\neq j \eqcm \nn\\
\tilde x_j=\frac{\bar x_j+\xi}{1+\xi} \eqcm &\qquad& 
\tilde{\bf k}_{\perp j}=\bar{\bf k}_{\perp j}
                        -\frac{1-\bar x_j}{1+\xi}\,
                         \frac{{\bf\Delta}_\perp}{2}
\eqpt
\end{eqnarray} 
Likewise, the LCWF arguments for the outgoing hadron (i.e., the
momenta of the partons belonging to the outgoing hadron in the
hadron-out frame) are related to the momenta in the average-frame by
\begin{eqnarray} 
\label{eq:hat-args}
\hat x'_i= \frac{\bar x_i}{1-\xi} \eqcm &\qquad& 
\hat{\bf k}'_{\perp i}=\bar{\bf k}^\adj_{\perp i}
                        -\frac{\bar x_i}{1-\xi}\,
                         \frac{{\bf\Delta}_\perp}{2} \eqcm
\qquad \qquad \mbox{for }i\neq j \eqcm \nn\\
\hat x'_j=\frac{\bar x_j-\xi}{1-\xi} \eqcm &\qquad& 
\hat{\bf k}'_{\perp j}=\bar{\bf k}^\adj_{\perp j}
                        +\frac{1-\bar x_j}{1-\xi}\,
                         \frac{{\bf\Delta}_\perp}{2}
\eqpt
\end{eqnarray} 
Using this we can express the single-particle state normalisation
(\ref{eq:statenorm}) through the LCWF arguments
\begin{eqnarray} 
\langle s'_i;w'_i|s_i;w_i\rangle&=&
16\pi^3\,\hat x'_i\,
\delta\left(\hat x'_i-\tilde x^\adj_i\,
\frac{1+\xi}{1-\xi}\right)\;
\nn\\
&\times&
\delta^{(2)}\! \left(\hat{\bf k}'_{\perp i}
-\tilde{\bf k}^\adj_{\perp i}+
\frac{\tilde x_i}{1-\xi}{\bf \Delta}^\adj_\perp\right)\;
\delta_{\mu_i'\mu^\adj_i} \;
\delta_{s'_i s_i^\adj}\;
\delta_{c'_i c_i^\adj}
\eqcm
\label{eq:spectatornorm}
\end{eqnarray} 
where the relations (\ref{eq:tilde-args}) and (\ref{eq:hat-args}) have
been used to express the variables $w_i$ and $w'_i$ in terms of hadron
frame quantities (with tilde and hat, respectively) and thus in terms
of the variables occurring in the integration measures.  Now we use
the expansion (\ref{eq:density-expansion}) of the density operator, of
which only the term with the quark operators $b^\dagger(w')\, b(w)$
contributes to the matrix element here. Combined with 
the use of the definition of the single-quark state (\ref{eq:partonstate}) 
and the anticommutation relation for the quark creation and annihilation
operators this yields
\begin{eqnarray} 
\lefteqn{
\sum_c
\int\frac{\d z^-}{2\pi}\;e^{i\,\bar x\,\bar p^{\,+}z^-} \,
\langle s'_j;w'_j|
        \,\phi^{\,c\,\dagger}_q(-\bar z/2)\,\phi^{\,c}_q(\bar z/2)\,
        |s^\adj_j;w^\adj_j\rangle}
\hspace{6em} \nn\\
&=&
\frac{1}{\bar p^{\,+}}\;
\delta(\bar x-\bar x_j)\;
u_+^\dagger(w'_j)\,u_+(w^\adj_j)\;
\delta_{s_j q}\; \delta_{s'{}\!\!_j s_j}\,
\delta_{c'{}\!\!_j c_j}
\eqpt
\label{eq:activematrix}
\end{eqnarray} 
The argument of the $\delta$-function is simplified with the help of
(\ref{eq:auxiliary}). Thus we arrive at
\begin{eqnarray}
{\cal H}_{\lambda'\lambda}^{q(N\to N)}&=& 
\frac{1}{\bar p^{\,+}\sqrt{2(1-\xi^2)}}\;
\sqrt{\frac{1-\xi}{1+\xi}}^{\,1-N}\;
\sum_{\beta,\beta'}\;\sum_{j=1}^N\;
\int [\d\tilde x]_N [\d^2 \tilde{\bf k}_\perp]_N\;
\d \hat x'_j \, \d^2 \hat{\bf k}'_{\perp j}
\nn\\
&\times&
\delta\!\left(\hat x'_j-1+\sum_{i\ne j}
 \tilde x^\adj_i\,\frac{1+\xi}{1-\xi}\right)\;
\delta^{(2)}\!\left(\hat{\bf k}'_{\perp j}
  +\sum_{i\ne j}\left(\tilde{\bf k}^\adj_{\perp i}-\tilde x^\adj_i\,
 \frac{1+\xi}{1-\xi}{\bf\Delta}_\perp^\adj\right)\right)\nn\\
&\times&
\delta(\bar x-\bar x_j)\;
u_+^\dagger(w'_j)\,u_+(w^\adj_j)\;
\delta_{s_j q}\; \delta_{s'{}\!\!_j s_j}
\delta_{c'{}\!\!_j c_j}
\prod_{i=1\atop i\neq j}^N 
\delta_{s'_i s^\adj_i}\;
\delta_{\mu_i'\mu^\adj_i}\;
\delta_{c'_i c^\adj_i}
\nn\\
&\times&
\frac{\Psi_{N,\beta'}^{*\,\lambda'}(\hat r')\,
      \Psi_{N,\beta}^\lambda(\tilde r)}{\sqrt{\hat x'_j\tilde x_j}}\;
\eqpt
\label{nflastbutone}
\end{eqnarray}
Note that the $\delta$-functions shown explicitly, together with the
ones in the integration measures $[\d\tilde x]_N [\d^2 \tilde{\bf
k}_\perp]_N$, provide the relation given in (\ref{eq:hat-args}) for
the active quark momentum variables $\hat x_j'$ and $\hat{\bf
k}'_{\perp j}$. The integrations over $\hat x_j'$ and $\hat{\bf
k}'_{\perp j}$ can be carried out, and in the following the primed
variables (with hat) are used as a shorthand defined by
(\ref{eq:hat-args}). The 
equality of flavours ($s'_j$ and $s_j$) and colours ($c'_j$ and
$c_j$) is assured by the Kronecker symbols in
Eq.~(\ref{nflastbutone}), and the spinor product evaluates to
\begin{equation} 
u_+^\dagger(w'_j)\,u_+(w_j)=
\frac{1}{\sqrt{2}}\;
\bar u(w'_j)\,\gamma^+\,u(w_j)=
\sqrt{2\,(1-\xi^2)\,\hat x'_j\,\tilde x_j}\;\bar p^{\,+} \;
\delta_{\mu_j'\mu^\adj_j}
\eqpt
\label{spinor-product}
\end{equation} 
The Kronecker-$\delta$ for the helicities in (\ref{spinor-product})
could have been anticipated from the Dirac structure of the operator
$\ovl\psi{}_q^{\,c}\gamma^+\psi_q^{\,c}$ in
(\ref{eq:quarkSPDdef}). Defining right- and left-handed projections of
the fields, $\phi_{q,R/L}^{\,c}\equiv P_{R/L}\,P_+\psi_q^{\,c}$, with
$P_{R/L}=(1\pm \gamma_5)/2$ and using $P_\pm\gamma_5=\gamma_5 P_\pm$
it is easy to see that
\begin{equation}
\ovl\psi{}_q^{\,c}\gamma^+\psi_q^{\,c}
=\sqrt{2}\;\phi_q^{\,c\,\dagger}(P_R+P_L)\,\phi_q^{\,c}
=\sqrt{2}\;\left(\phi_{q\,R}^{\,c\,\dagger}\phi_{q\,R}^{\,c}
                +\phi_{q\,L}^{\,c\,\dagger}\phi_{q\,L}^{\,c}\right)
\label{eq:density-sum}
\eqpt
\end{equation} 
Since for massless quarks chirality and helicity are identical, the
helicities on both quark lines have to be the same.

To present our final result in a symmetric way we rewrite the
integration measure in terms of the average quantities with the help
of (\ref{eq:hat-args})
\begin{equation} 
[\d\tilde x]_N=
\left(\frac{1}{1+\xi}\right)^{N-1}\,[\d\bar x]_N 
\eqcm\qquad\qquad
[\d^2 \tilde{\bf k}_\perp]_N=[\d^2 \bar{\bf k}_\perp]_N
\eqcm
\end{equation}
and arrive at the overlap representation of the quark SPD in the
region $\xi<\bar x<1$:
\begin{eqnarray} 
{\cal H}_{\lambda'\lambda}^{q(N\to N)}&=&
\sqrt{1-\xi}^{\,1-N}\sqrt{1+\xi}^{\,1-N} \;
\sum_{\beta=\beta'}\;
\sum_{j} \delta_{s_j q}\;
\nn\\
&\times&
\int [\d\bar x]_N [\d^2 \bar{\bf k}_\perp]_N \;
\delta\left(\bar x-\bar x_j\right)\;
\Psi_{N,\beta'}^{*\,\lambda'}(\hat r')\,
\Psi_{N,\beta}^\lambda(\tilde r)
\eqcm
\label{eq:quarkSPD}
\end{eqnarray} 
with the arguments $\tilde r$ ($\hat r'$) of the LCWF for the incoming
(outgoing) proton being related to the integration variables $\bar x_i$
and $\bar {\bf k}_{\perp i}$ by (\ref{eq:tilde-args}) and
(\ref{eq:hat-args}), respectively. Summation over $N$ leads to the 
full expression of
${\cal H}_{\lambda'\lambda}^q$ in the region $\xi<\bar x<1$.
Alternatively, as we discussed in Sect.~\ref{sec:fock}, one could use
$N$-parton Fock states that are coupled to be colourless and carry the
quantum numbers of the hadron. Normalising these states in analogy to
Eq.~(\ref{eq:parton-state-norm}) and denoting the associated LCWFs by
$\widetilde{\Psi}_{N,\tilde{\beta}}$, one would obtain an overlap
$\widetilde{\Psi}^{*}{}_{\!\!\! N,\tilde{\beta'}}\,
\widetilde{\Psi}_{N,\tilde{\beta}}$ summed over
$\tilde{\beta}=\tilde{\beta}'$, just as in (\ref{eq:quarkSPD}). This
simple structure is owed to the constraint $\beta=\beta'$ in
(\ref{eq:quarkSPD}) and will no longer appear in the region $-\xi<\bar
x<\xi$, see Eq.~(\ref{eq:quarkSPD-ERBL}).

%%%%%%%%%%%%%%%%%%%%%%%%%%%%%%%%%%%
\subsection{The region $-1<\bar x<-\xi$}

For active {\em antiquarks}\ the derivation of the overlap
representation of the SPDs goes in full analogy to the one we have
just given. Differences appear when the Fourier decomposition
(\ref{eq:density-expansion}) is used to yield the analogue of
Eq.~(\ref{eq:activematrix}), since it is now the term with $d(w')\,
d^\dagger(w)$ that contributes. Exchanging the order of the
annihilation and creation operator gives an overall minus sign, and
the $\delta$-function in the analogue of Eq.~(\ref{eq:activematrix}) now
gives the constraint $\bar x_j=-\bar x$. The final result for the
region $-1<\bar x<-\xi$ is
\begin{eqnarray} 
{\cal H}_{\lambda'\lambda}^{q(N\to N)}&=&
{}- \sqrt{1-\xi}^{\,1-N}\sqrt{1+\xi}^{\,1-N} \;
\sum_{\beta=\beta'}\;
\sum_{j} \delta_{\bar s_j q}\;
\nn\\
&\times&
\int [\d\bar x]_N [\d^2 \bar{\bf k}_\perp]_N \;
\delta\left(\bar x+\bar x_j\right)\;
\Psi_{N,\beta'}^{*\,\lambda'}(\hat r')\,
\Psi_{N,\beta}^\lambda(\tilde r)
\eqcm
\label{eq:antiquarkSPD}
\end{eqnarray} 
with the LCWF arguments $\tilde r$ and $\hat r'$ given by
(\ref{eq:tilde-args}) and (\ref{eq:hat-args}), respectively

%%%%%%%%%%%%%%%%%%%%%%%%%%%%%%%%%%%%%%%%%
\subsection{The region $-\xi<\bar x<\xi$}
\label{sec:central}

Let us now consider the kinematical range $-\xi<\bar x<\xi$. As
mentioned above, we restrict ourselves to the case $\xi>0$. Therefore,
the quark SPDs in this region describe the emission of a
quark-antiquark pair from the initial proton. In the Fock state
decompositions of the initial and final protons we thus have to
consider only terms where the initial state has the same parton
content as the final state plus one additional quark-antiquark
pair. We thus have
\begin{equation}
\label{eq:Focksummation-ERBL}
{\cal H}_{\lambda'\lambda}^q=
\sum_N {\cal H}_{\lambda'\lambda}^{q(N+1\to N-1)}
\end{equation}  
as opposed to (\ref{eq:Focksummation}). This particular type of
overlap was recently identified in \cite{Brodsky:1999hn} in the
context of transition form factors between heavy and light mesons.

Starting from the definition (\ref{eq:quarkSPDdef}) of SPDs for quarks
of flavour $q$ and replacing the hadronic states by their Fock state
decomposition (\ref{eq:Fockstate}), the contribution of the $N+1\to
N-1$ transition to the matrix element ${\cal H}_{\lambda'\lambda}^q$
is found to be
\begin{eqnarray}
\lefteqn{
{\cal H}_{\lambda'\lambda}^{q(N+1\to N-1)} =
\frac{1}{\sqrt{2(1-\xi^2)}}\;
\sum_c
\sum_{\beta,\beta'}\;
\int [\d \tilde x]_{N+1} [\d^2 \tilde{\bf k}_\perp]_{N+1} \;
     [\d \hat x']_{N-1} [\d^2 \hat{\bf k}'_\perp]_{N-1} \;
} \nn \\
&\times&
\Psi_{N-1,\beta'}^{*\,\lambda'}(\hat r')\,
      \Psi_{N+1,\beta}^\lambda(\tilde r)
\int\frac{\d z^-}{2\pi}\;e^{i\,\bar x\,\bar p^{\,+}z^-}
\nn\\[\baselineskip]
&\times&
\left\langle N-1,\beta';k'_1\ldots k'_{N-1}\right|
        \,\phi^{\,c\,\dagger}_{q}(-\bar z/2)\,\phi_{q}^{\,c}(\bar z/2)\,
              \left|\phantom{k'_{N-1}}\hspace{-4.6ex}
                    N+1,\beta;k_1,\ldots,k_{N+1}\right\rangle\;
\eqpt \hspace{2em}
\label{ERBL-start}
\end{eqnarray}
Using again the (anti)commutation relations for the creation and annihilation
operators the partonic matrix element can be replaced by
\begin{eqnarray}
\lefteqn{
\left\langle N-1,\beta';k'_1\ldots k'_{N-1}\right|
        \,\phi^{\,c\,\dagger}_{q}(-\bar z/2)\,\phi_{q}^{\,c}(\bar z/2)\,
      \left|\phantom{k'_{N-1}}\hspace{-4.6ex}
                    N+1,\beta;k_1,\ldots,k_{N+1}\right\rangle\;}
\nn\\[\baselineskip]
&=&
\sum_{j,j'=1}^{N+1}\;
\frac{1}{\sqrt{n_j n_{j'} \rule{0pt}{0.7em}}}\;
\frac{\langle 0|
        \,\phi^{\,c\,\dagger}_{q}(-\bar z/2)\,\phi_{q}^{\,c}(\bar z/2)\,
                            |s_j,w_j; s_{j'},w_{j'}\rangle}
{\sqrt{\hat x'_1\ldots \hat x'_{N-1}}
   \sqrt{\tilde x_1\ldots \tilde x_{N+1}\phantom{\hat x'_{N-1}}
   \hspace{-4.4ex}}
   }\;
\prod_{i=1\atop i\ne j,j'}^{N+1} 
\langle s'_i;w'_i|s^\adj_i;w^\adj_i\rangle
\eqcm \hspace{2em}
\label{eq:ERBL-activematrix}
\end{eqnarray}
where we label the active quark-antiquark pair with indices $j$ for
the quark and $j'$ for the antiquark, and write 
$|s_j,w_j; s_{j'},w_{j'}\rangle=
 b^\dagger_{s_j}(w_j)\,d^\dagger_{s_{j'}}(w_{j'})\,|0\rangle$ for the 
corresponding two-parton state. The sum over $\beta$ and $\beta'$ now 
runs over combinations  where flavour, helicity (and colour) of all 
spectators match. $n_j$ ($n_{j'}$) is the number of
(anti)quarks in the initial proton wave function
$\Psi_{N+1,\beta}^\lambda(r)$ with the same discrete quantum numbers
as the active (anti)quark. These factors appear since the product
$\sqrt{f_{N+1,\beta}\, f_{N-1,\beta'}}$ of normalisation factors from
the parton states (\ref{parton-states}) is not equal to the number of
possibilities to associate the partons in the initial and final proton,
in contrast to the situation in the regions discussed so far.

To simplify the notation we use the same numbering for the spectator
partons in the LCWFs of the initial and final state proton. Thus the
$N-1$ partons in the outgoing proton are numbered not as $i=1, \ldots,
N-1$, but as $i=1, \ldots, N+1$ with $j$ and $j'$ omitted. From the
spectator momenta $k^\adj_i$ and $k'_i$ ($i\neq j,j'$) we again form
the auxiliary variables defined in Eq.~(\ref{eq:auxiliary}). For $j$
and $j'$ we introduce
\begin{equation}
\bar k_j=\frac{1}{2}\, (k_j-k_{j'}) \;,\qquad
\bar x_j=\frac{\bar k_j^+}{\bar p^{+}}
\eqcm
\end{equation} 
which is half the relative momentum (and momentum fraction) between
the active quark and antiquark. It can as well be viewed as the
average of $k_j$ and the reversed momentum $k_{j'}$ (i.e., $-k_{j'}$),
in complete analogy with the definitions (\ref{eq:auxiliary}). {}From
momentum conservation and the spectator condition
\begin{equation} 
k'_i=\bar k_i^\adj=k_i^\adj \eqcm
\qquad \mbox{for }i\neq j,j' 
\end{equation} 
we now obtain that the LCWF arguments for the incoming hadron are
related to the parton momenta in the average-frame by
\begin{eqnarray}
\label{eq:tilde-args-central}
\tilde x_i= \frac{\bar x_i}{1+\xi} \eqcm &\qquad& 
\tilde{\bf k}_{\perp i}=\bar{\bf k}_{\perp i}
                        +\frac{\bar x_i}{1+\xi}\,
                         \frac{{\bf\Delta}^\adj_\perp}{2} \eqcm
\qquad \qquad \mbox{for }i\neq j,j' 
\eqcm \nn\\
\tilde x_j=\frac{\bar x_j+\xi}{1+\xi} \eqcm &\qquad& 
\tilde{\bf k}_{\perp j}=\bar{\bf k}_{\perp j}
                        -\frac{1-\bar x_j}{1+\xi}\,
                         \frac{{\bf\Delta}^\adj_\perp}{2}
\eqcm \nn\\
\tilde x_{j'}=-\frac{\bar x_j-\xi}{1+\xi} \eqcm &\qquad& 
\tilde{\bf k}_{\perp j'}=-\bar{\bf k}_{\perp j}
                        -\frac{1+\bar x_j}{1+\xi}\,
                         \frac{{\bf\Delta}^\adj_\perp}{2}
\eqcm
\end{eqnarray}
and that the LCWF arguments for the outgoing hadron are given by
\begin{eqnarray} 
\label{eq:hat-args-central}
\hat x'_i= \frac{\bar x_i}{1-\xi} \eqcm &\qquad& 
\hat{\bf k}'_{\perp i}=\bar{\bf k}^\adj_{\perp i}
                        -\frac{\bar x_i}{1-\xi}\,
                         \frac{{\bf\Delta}^\adj_\perp}{2}
\eqcm
\qquad \qquad \mbox{for }i\neq j,j' 
\eqpt
\end{eqnarray} 
The relations (\ref{eq:tilde-args-central}) and
(\ref{eq:hat-args-central}) can be used to write $\langle
s';w'|s;w\rangle$ again as in Eq.~(\ref{eq:spectatornorm}), and for
the evaluation of the matrix element of the active quark-antiquark
pair we insert the expansion (\ref{eq:density-expansion}) of the
density operator, now keeping only the term with $d(w')\, b(w)$. 
We then use the definition of the two-parton state given after 
(\ref{eq:ERBL-activematrix}), and the anticommutation
relations (\ref{eq:anticommutation}) to write
\begin{eqnarray}
\lefteqn{
\sum_c
\int\frac{\d z^-}{2\pi}\;e^{i\,\bar x\,\bar p^{\,+}z^-} \,
\langle 0|
        \,\phi^{\,c\,\dagger}_q(-\bar z/2)\,\phi^{\,c}_q(\bar z/2)\,
        |s_j,w_j; s_{j'},w_{j'}\rangle}
\hspace{6em} \nn\\
&=&
\frac{1}{\bar p^{\,+}}\;
\delta(\bar x-\bar x_j)\;
v_+^\dagger(w_{j'})\,u_+(w^\adj_j)\;
\delta_{{\bar s}_{j'}s^\adj_j}\;\delta_{s^\adj_j q}\;
\delta_{{c}_{j'}c^\adj_j}
\eqpt
\label{eq:activematrix-ERBL}
\end{eqnarray} 
Putting the pieces together we arrive at
\begin{eqnarray}
\label{eq:quarkSPDstep}
\lefteqn{
{\cal H}_{\lambda'\lambda}^{q(N+1\to N-1)}= 
\frac{1}{\bar p^{\,+}\sqrt{2(1-\xi^2)}}\;
\sqrt{\frac{1-\xi}{1+\xi}}^{\,1-N}\;
\sum_{\beta,\beta'}\;
\sum_{j,j'=1}^{N+1}
\frac{1}{\sqrt{n_j n_{j'} \rule{0pt}{0.7em}}}
}
\nn\\
&\times&
\int [\d\tilde x]_{N+1} [\d^2 \tilde{\bf k}_\perp]_{N+1}\,
16\pi^3\;
\delta\!\left(1-\frac{1+\xi}{1-\xi}
              \sum_{i=1\atop i\ne j,j'}^{N+1} \tilde x_i\right)\;
\delta^{(2)}\!\left(
  \frac{{\bf\Delta}_\perp}{1+\xi} -
  \sum_{i=1\atop i\ne j,j'}^{N+1}\tilde{\bf k}_{\perp i}
  \right)
\nn\\
&\times&
\delta(\bar x-\bar x_j)\;
v_+^\dagger(w_{j'})\,u_+(w_j)\;
\delta_{{\bar s}_{j'}s^\adj_j}\;\delta_{s^\adj_j q}\;
\delta_{{c}_{j'}c^\adj_j}\;
\prod_{i=1\atop i\ne j,j'}^{N+1} 
\delta_{s'_i s^\adj_i}\;
\delta_{\mu_i'\mu^\adj_i}\;
\delta_{c'_i c^\adj_i}
\nn\\
&\times&
\frac{\Psi_{N-1,\beta'}^{*\,\lambda'}(\hat r')\,
      \Psi_{N+1,\beta}^\lambda(\tilde r)}{
                                      \sqrt{\tilde x_j\tilde x_{j'}}}
\eqpt \hspace{2em}
\end{eqnarray}
The spinor product occurring is
\begin{equation}
v_+^\dagger(w_{j'})\,u_+(w_j)=
\frac{1}{\sqrt{2}}\;
\bar v(w_{j'})\,\gamma^+\,u(w_j)=
\sqrt{2\,\tilde x_j\,\tilde x_{j'}}\;(1+\xi)\;\bar p^{\,+} \;
\delta_{\mu_{j'}-\mu^\adj_j}
\eqpt
\end{equation}
The integrations over $\tilde x_{j'}$ and $\tilde{\bf k}_{\perp j'}$
can be carried out, and by rewriting the remaining integrations in
terms of the auxiliary variables we arrive at the overlap
representation of ${\cal H}_{\lambda'\lambda}^q$ in the region
$-\xi<\bar x<\xi$ for the $N+1\to N-1$ transition:
\begin{eqnarray}
\label{eq:quarkSPD-ERBL}
{\cal H}_{\lambda'\lambda}^{q(N+1\to N-1)} &=&
\sqrt{1-\xi}^{\,2-N} \sqrt{1+\xi}^{\,-N} \;
\sum_{\beta,\beta'}\;
\sum_{j,j'=1}^{N+1}
\frac{1}{\sqrt{n_j n_{j'} \rule{0pt}{0.7em}}}\;
\delta_{{\bar s}_{j'}s^\adj_j}\;\delta_{s^\adj_j q}\;
\delta_{\mu_{j'}-\mu^\adj_j}\,
\delta_{{c}_{j'}c^\adj_j}\;
\nn\\
&\times&
\prod_{i=1\atop i\ne j,j'}^{N+1} 
\delta_{\mu_i'\mu^\adj_i}\;
\delta_{s'_i s^\adj_i}\;
\delta_{c'_i c^\adj_i}\;
\int \d\bar x_j
\prod_{i=1\atop i\ne j,j'}^{N+1} 
\d\bar x_i\;
\delta\left(1-\xi-\sum_{i=1\atop i\ne j,j'}^{N+1}\bar x_i\right)\;
\nn\\
&\times&
\int \d^2 \bar{\bf k}_{\perp j}
\prod_{i=1\atop i\ne j,j'}^{N+1} \d^2 \bar{\bf k}_{\perp i} \;
(16\pi^3)^{1-N}\;
\delta^{(2)}\!\left(\frac{{\bf\Delta}_\perp}{2}
              - \sum_{i=1\atop i\ne j,j'}^{N+1} \bar{\bf k}_{\perp i}
              \right)
\nn\\
&\times&
\delta\left(\bar x-\bar x_j\right)\; 
\Psi_{N-1,\beta'}^{*\,\lambda'}(\hat r')\,
\Psi_{N+1,\beta}^\lambda(\tilde r)
\eqcm \phantom{ \prod_{i=1\atop i\ne j,j'}^{N+1} }
\end{eqnarray}
The arguments
$\tilde r$ and $\hat r'$ of the wave functions are given in terms of
$\bar x_i$ and $\bar {\bf k}_{\perp i}$ by
(\ref{eq:tilde-args-central}) and (\ref{eq:hat-args-central}), and
$n_j$, $n_{j'}$ are defined after Eq.~(\ref{ERBL-start}).
As was to be expected, the operator $\sum_c \bar\psi^{\,c}_q \gamma^+
\psi_q^{\,c}$ in (\ref{eq:quarkSPDdef}) projects out colour singlet $q
\bar{q}$ pairs with total helicity zero in the initial proton LCWF.

At this point we see the advantage of our method to start from the
parton states (\ref{parton-states}). If one uses colour neutral parton
states with the quantum numbers of the hadron under investigation, one
has in general to rearrange their colour coupling to ensure that the
active $q \bar{q}$ pair is in a colour singlet state, and a
combinatorial factor different from $1/\sqrt{\rule{0pt}{1.4ex} n_j
n_{j'}}$ will then appear in (\ref{eq:quarkSPD-ERBL}). Although such a
procedure should be possible using appropriate group theoretical
methods (see e.g.\ \cite{Hofestadt:1987wk}), we have not pursued this
point here.

%%%%%%%%%%%%%%%%%%%%%%%%%%%%%%%%%%%%%%%%%%%%%%%%%%%%%%%%%%%%%%%%%%%%%%%%%
\section{Quark polarisation and gluons}
\label{sec:polglue}
%%%%%%%%%%%%%%%%%%%%%%%%%%%%%%%%%%%%%%%%%%%%%%%%%%%%%%%%%%%%%%%%%%%%%%%%%

%%%%%%%%%%%%%%%%%%%%%%%%%%%%%%%%%%%%%%%%%%%%%%%%
\subsection{The polarised skewed quark distribution}

We now turn to the polarised skewed quark distributions,
$\widetilde{H}^q(\bar x,\xi;t)$ and $\widetilde{E}^q(\bar x,\xi;t)$,
defined by the Fourier transform of the axial vector matrix element
\begin{eqnarray} 
\widetilde{\cal H}_{\lambda'\lambda}^q &\equiv&
\frac{1}{2\,\sqrt{1-\xi^2}} \;
\sum_c
\int\frac{\d z^-}{2\pi}\;e^{i\,\bar x\,\bar p^{\,+}z^-}\;
\langle p^{\,\prime},\lambda'|
   \,\ovl\psi{}_q^{\,c}(-\bar z/2)\,
     \gamma^+\gamma_5\,\psi_q^{\,c}(\bar z/2)\, |p,\lambda\rangle
\nn\\[1\baselineskip]
&=&
 \frac{\ovl u(p^{\,\prime},\lambda')\gamma^+\gamma_5 u(p,\lambda)}
      {2\bar p^{\,+}\sqrt{1-\xi^2}}\;
\widetilde{H}^q(\bar x,\xi;t)
+\frac{\ovl u(p^{\,\prime},\lambda')\Delta^+\gamma_5 u(p,\lambda)}
      {4m\,\bar p^{\,+}\sqrt{1-\xi^2}}\;
\widetilde{E}^q(\bar x,\xi;t)
\eqpt
\end{eqnarray} 
For the different proton helicity combinations we now find
\begin{eqnarray}
\label{eq:pol-quark-helcomb}
\widetilde{\cal H}^q_{++} = 
- \phantom{( } \widetilde{\cal H}^q_{--} \phantom{ )^*} &=& 
\widetilde{H}^q - \frac{\xi^2}{1-\xi^2}\, \widetilde{E}^q \eqcm
\nn\\
\widetilde{\cal H}^q_{-+} = 
\phantom{-} (\widetilde{\cal H}^q_{+-})^* &=& 
      \eta\, \frac{\sqrt{t_0 - t}}{2m}\, \frac{\xi}{\sqrt{1-\xi^2}}\, 
      \widetilde{E}^q \eqpt
\end{eqnarray}
The derivation of an overlap formula goes along the same lines as for
the unpolarised quark SPDs. We just need the appropriate conversion of
the quark field operators into a density of LC fields. Expressing the
axial vector operator in terms of the left- and right-handed
projections we obtain
\begin{equation} 
\label{eq:quark-hel-sel}
\ovl\psi{}_q^{\,c}\gamma^+\gamma_5\psi^{\,c}_q
=\sqrt{2}\;\phi^{\,c\,\dagger}_q\gamma_5\phi^{\,c}_q
=\sqrt{2}\;\left(\phi^{\,c\,\dagger}_{q\,R}\phi^{\,c}_{q\,R}
                -\phi^{\,c\,\dagger}_{q\,L}\phi^{\,c}_{q\,L}\right)
\eqpt
\end{equation}
Compared to (\ref{eq:density-sum}) the difference of density operators
for left- and right-handed projections now appears. Repeating all
steps in the derivation of (\ref{eq:quarkSPD}) one finds the overlap
representation of the contribution of the $N$ particle Fock state to
the SPD for a polarised quark of flavour $q$ in the region $\xi<\bar
x<1$
\begin{eqnarray} 
\label{eq:pol-quarkSPD}
\widetilde{\cal H}_{\lambda'\lambda}^{q(N\to N)}&=&
\sqrt{1-\xi}^{\,1-N}\sqrt{1+\xi}^{\,1-N} \;
\sum_{\beta=\beta'}\;
\sum_{j}
\mbox{sign}(\mu_j)\;\delta_{s_j q}\;
\nn\\
&\times&
\int [\d\bar x]_N [\d^2 \bar{\bf k}_\perp]_N \;
\delta\left(\bar x-\bar x_j\right)\;
\Psi_{N,\beta'}^{*\,\lambda'}(\hat r')\,
\Psi_{N,\beta}^\lambda(\tilde r)
\eqcm
\end{eqnarray} 
where the arguments of the LCWFs are again given by the relations
(\ref{eq:tilde-args}) and (\ref{eq:hat-args}). The only difference
between the RHS of (\ref{eq:pol-quarkSPD}) and the RHS of
(\ref{eq:quarkSPD}) is the sign-function of the helicity of the active
quark. Summation over all Fock states as in (\ref{eq:Focksummation})
leads to the final result.  In the region $-1<\bar{x}<-\xi$ one has
\begin{eqnarray} 
\label{eq:pol-antiquarkSPD}
\widetilde{\cal H}_{\lambda'\lambda}^{q(N\to N)}&=&
\sqrt{1-\xi}^{\,1-N}\sqrt{1+\xi}^{\,1-N} \;
\sum_{\beta=\beta'}\;
\sum_{j}
\mbox{sign}(\mu_j)\;\delta_{\bar s_j q}\;
\nn\\
&\times&
\int [\d\bar x]_N [\d^2 \bar{\bf k}_\perp]_N \;
\delta\left(\bar x+\bar x_j\right)\;
\Psi_{N,\beta'}^{*\,\lambda'}(\hat r')\,
\Psi_{N,\beta}^\lambda(\tilde r)
\eqpt
\end{eqnarray} 
In contrast to Eq.~(\ref{eq:antiquarkSPD}) there is no global minus
sign here, because $\mbox{sign}(\mu_j)$ refers to the antiquark
helicity, which is minus the chirality selected by the operator
(\ref{eq:quark-hel-sel}). The non-diagonal overlap $\widetilde{\cal
H}_{\lambda'\lambda}^{q(N+1\to N-1)}$ in the central region is
identical to (\ref{eq:quarkSPD-ERBL}) except for an additional factor
$\mbox{sign}(\mu_j)$, as appears in (\ref{eq:pol-quarkSPD}), where
$\mu_j$ refers to the helicity of the active quark.

%%%%%%%%%%%%%%%%%%%%%%%%%%%%%%%%%%%%%%
\subsection{The unpolarised skewed gluon distribution}
\label{sec:gluonSPD}

The unpolarised skewed gluon distributions $H^g(\bar x,\xi;t)$ and
$E^g(\bar x,\xi;t)$ are defined from the Fourier transform of a
hadronic matrix element involving two gluon field strength tensors at
a light-like distance:
\begin{eqnarray}
\label{eq:gSPDdef-ji}
{\cal H}^g_{\lambda'\lambda} &\equiv& 
\frac{-g_{\perp\, \alpha'\alpha}}{\bar p^{\,+}\sqrt{1-\xi^2}}\;
\sum_c
\int \frac{\d z^-}{2\pi} e^{i\bar x\,\bar p^{\,+}z^-} \;
\langle p',\lambda'|
        \,G_c^{+\alpha'}(-\bar z/2)\,G_c^{+\alpha}(\bar z/2)\,
                                  |p,\lambda\rangle
\nn\\[1\baselineskip]
&=&
 \frac{\ovl u(p',\lambda')\gamma^+ u(p,\lambda)}
      {2\bar p^{\,+}\sqrt{1-\xi^2}}\;
H^g(\bar x,\xi;t)
+\frac{\ovl u(p',\lambda')i\sigma^{+\alpha}\Delta_\alpha u(p,\lambda)}
      {4m\,\bar p^{\,+}\sqrt{1-\xi^2}}\;
E^g(\bar x,\xi;t)
\end{eqnarray}
with the transverse metric tensor $g_{\perp}^{\alpha'\alpha}$, which
has $g_{\perp}^{11}=g_{\perp}^{22}=-1$ as only non-zero
elements. Again the link operator is not displayed. Notice that in the
$A^+=0$ gauge the relation $G_c^{+\alpha}=\partial^+ A_c^\alpha$
provides a simple transition from field strengths to potentials. The
normalisation in Eq.~(\ref{eq:gSPDdef-ji}) is chosen such that in the
forward limit the SPD $H^g$ is related to the ordinary gluon
distribution $g(x)$ (for the definition see~\cite{Brock:1995sz}) by
\begin{equation}
\label{eq:forwardgluon}
H^g(\bar x,\xi=0;t=0)=\bar x\,g(\bar x)
\eqpt
\end{equation} 
The derivation of an overlap formula proceeds in close analogy to the
one for the quark SPDs. There is only one technical point we have to
comment on. {}From the expansion (\ref{eq:gluonexpansion}) of the
transverse components of the gluon field operators in momentum space,
one encounters a combination of polarisation vectors for the active
gluons, which in the region $\xi<\bar{x}<1$ reads
\begin{equation}
\epsilon^{*\,\alpha'}(w'_j)\,\epsilon^{\alpha}(w_j) \eqcm
\end{equation} 
and is to be contracted with $g_{\perp\,\alpha'\alpha}$. In a frame
where the active on-shell gluon in the incoming hadron has no
transverse momentum, its polarisation vector is purely transverse and
does not depend on its momentum:
\begin{equation} 
\epsilon(k^+_j,{\bf k}_{\perp j}=0,\mu_j)=
 \lcvec{0}{0}{{\vec\epsilon}_\perp(\mu_j)}
\eqcm
\end{equation} 
where ${\vec\epsilon}_\perp(\mu)=(-\mu,-i)/\sqrt{2}$. A transverse
boost from this frame to the hadron-in frame leaves the transverse
components unchanged. This is because the plus component of the
polarisation vector in the starting frame is zero. The plus component
itself remains zero by virtue of the definition of a transverse
boost. The transformation produces a non-zero minus component which we
do not have to specify, since the polarisation vector will be
contracted with a transverse tensor. A similar argument can be given
for the polarisation vector of the reabsorbed gluon, starting in a
frame where its momentum has no transverse component and applying an
appropriate transverse boost. Explicitly, we get
\begin{eqnarray} 
\epsilon^{*\,\alpha'}(w'_j)\, \epsilon^\alpha(w_j)&=&
-\frac{1}{2}
\left(g_{\perp}^{\alpha'\alpha}
  -\mbox{sign}(\mu_j)\,i\,\varepsilon_{\perp}^{\alpha'\alpha}\right)\,
\delta_{\mu'_j\mu^\adj_j}
- \frac{1}{2}\;
t_{\perp}^{\,\alpha'\alpha}\;
\delta_{\mu'_j-\mu^\adj_j}
\nn\\
&& + \mbox{``non-transverse''} \rule{0pt}{3ex}
\eqcm
\label{eq:polarisation-matrix}
\end{eqnarray} 
where $\varepsilon_{\perp}^{12}=-\varepsilon_{\perp}^{21}=1$,
$t_{\perp}^{\,11}=-t_{\perp}^{\,22}=1$, and
$t_{\perp}^{\,12}=t_{\perp}^{\,21}=i\mu_j$, while all other components
of these tensors are zero. The term ``non-transverse'' stands for a
matrix with vanishing matrix elements in the transverse
sub-space. Neither this matrix nor $t_{\perp}^{\alpha'\alpha}$
contribute when contracted with transverse tensors
$g_{\perp\,\alpha'\alpha}$ and $\varepsilon_{\perp\,\alpha'\alpha}$,
which occur in the definition of the unpolarised and polarised skewed
gluon distributions.
 
The final result for the overlap representation for the unpolarised
gluon SPD in the region $\xi<\bar x<1 $ is
\begin{eqnarray} 
\label{eq:gluonSPD}
{\cal H}_{\lambda'\lambda}^{g(N\to N)}&=&
\sqrt{\bar x^2-\xi^2}\;
\sqrt{1-\xi}^{\,1-N}\sqrt{1+\xi}^{\,1-N}\; 
\sum_{\beta=\beta'}\;
\sum_{j}
\delta_{s_j g}\;
\nn\\
&\times&
\int [\d\bar x]_N [\d^2 \bar{\bf k}_\perp]_N \;
\delta\left(\bar x-\bar x_j\right)\;
\Psi_{N,\beta'}^{*\,\lambda'}(\hat r')\,
\Psi_{N,\beta}^\lambda(\tilde r)\,
\eqcm
\end{eqnarray}
where the sum over $j$ runs over all gluons, and the arguments,
$\tilde r$ and $\hat r'$ of the LCWFs are related to the auxiliary
variables $\bar x_i$ and $\bar {\bf k}_{\perp i}$ by the relations
(\ref{eq:tilde-args}) and (\ref{eq:hat-args}).

For the region $-1<\bar x<-\xi$ the gluon SPDs can be readily obtained
from (\ref{eq:gluonSPD}) with the observation that $H^g$ and $E^g$ are
even functions in $\bar x$, since the gluon is its own antiparticle.

In the region $-\xi<\bar x<\xi$ the gluon SPDs describe the emission
of two gluons from the initial proton. In the Fock state
decomposition, therefore, we have to consider $N+1\to N-1$
transitions, where the initial state has the same parton content as
the final state plus two additional gluons. In the derivation of an
overlap representation one encounters a combination of polarisation
vectors, which can be evaluated exactly along the lines discussed
above as
\begin{eqnarray} 
\label{eq:polarisation-matrix-ERBL}
\epsilon^{\alpha'}(w_{j'})\, \epsilon^\alpha(w_j)&=&
 \frac{1}{2}\,
\left(g_{\perp}^{\alpha'\alpha}
  -\mbox{sign}(\mu_j)\,i\,\varepsilon_{\perp}^{\alpha'\alpha}\right)\,
\delta_{\mu_{j'}-\mu^\adj_j}
+ \frac{1}{2}\; t_{\perp}^{\,\alpha'\alpha}\;
\delta_{\mu_{j'}\mu^\adj_j}
\nn\\
&&
+ \mbox{``non-transverse''} \rule{0pt}{3ex}
\eqpt
\end{eqnarray} 
Accordingly, the overlap representation of ${\cal
H}^g_{\lambda'\lambda}$ in the region $-\xi<\bar x<\xi$ for the
$N+1\to N-1$ transition becomes
\begin{eqnarray}
{\cal H}_{\lambda'\lambda}^{g(N+1\to N-1)} &=& 
- \sqrt{\xi^2-\bar x^2}\;
\sqrt{1-\xi}^{\,2-N} \sqrt{1+\xi}^{\,-N} 
\nn\\
&\times&
\sum_{\beta,\beta'}\;
\sum_{j,j'=1 \atop j\neq j'}^{N+1}
\frac{1}{\sqrt{n_j n_{j'} \rule{0pt}{0.7em}}}\;
\delta_{s_{j'}s_j}\;\delta_{s_j g}\;
\delta_{\mu_{j'}-\mu^\adj_j}\,
\delta_{\bar c_{j'}c_j}\;
\prod_{i=1\atop i\ne j,j'}^{N+1} 
\delta_{s'_i s^\adj_i}\;
\delta_{\mu_i'\mu^\adj_i}\;
\delta_{c'_i c^\adj_i}\;
\nn\\
&\times&
\int \d\bar x_j
\prod_{i=1\atop i\ne j,j'}^{N+1} 
\d\bar x_i\;
\delta\left(\sum_{i=1\atop i\ne j,j'}^{N+1}\bar x_i-(1-\xi)\right)\;
\nn\\
&\times&
\int \d^2 \bar{\bf k}_{\perp j}
\prod_{i=1\atop i\ne j,j'}^{N+1} \d^2 \bar{\bf k}_{\perp i} \;
(16\pi^3)^{1-N}\;
\delta^{(2)}\!\left(\sum_{i=1\atop i\ne j,j'}^{N+1}
            \bar{\bf k}_{\perp i}-\frac{{\bf\Delta}_\perp}{2}\right)
\nn\\
&\times&
\delta\left(\bar x-\bar x_j\right)\;
\Psi_{N-1,\beta'}^{*\,\lambda'}(\hat r')\,
\Psi_{N+1,\beta}^\lambda(\tilde r)
\eqcm \phantom{ \prod_{i=1\atop i\ne j,j'}^{N+1} }
\label{eq:gluonSPD-ERBL}
\end{eqnarray}
where both $j$ and $j'$ run over all gluons. The sum over $\beta$ and
$\beta'$ runs over combinations where flavour and colour of all
spectators match. The arguments $\tilde r$ and $\hat r'$ of the wave
functions are given in terms of $\bar x_i$ and $\bar {\bf k}_{\perp
i}$ by (\ref{eq:tilde-args-central}) and (\ref{eq:hat-args-central}).

We finally remark that compared with the quark SPDs, the overlap
representation for the case of gluons has an extra factor of
$\sqrt{|\bar x^2 - \xi^2|}$ in all regions of $\bar x$.

%%%%%%%%%%%%%%%%%%%%%%%%%%%%%%%%%%%%%%%%%%%%%%%%%%%%
\subsection{The polarised skewed gluon distribution}

The polarised gluon SPDs $\widetilde{H}^g(\bar x,\xi;t)$ and 
$\widetilde{E}^g(\bar x,\xi;t)$ are defined from
\begin{eqnarray} 
\label{eq:pol-gluonSPD} 
\widetilde{\cal H}^g_{\lambda'\lambda} &\equiv&
\frac{i\,\varepsilon^\adj_{\perp\,\alpha'\alpha}}
     {\bar p^{\,+}\sqrt{1-\xi^2}}\;
\sum_c
\int \frac{\d z^-}{2\pi} e^{i\bar x\,\bar p^{\,+}z^-} \;
\langle p^{\,\prime},\lambda'|
        \,G_c^{+\alpha'}(-\bar z/2)\,G_c^{+\alpha}(\bar z/2)\,
             |p,\lambda\rangle
\nn\\[1\baselineskip]
&=&
 \frac{\ovl u(p^{\,\prime},\lambda')\gamma^+\gamma_5 u(p,\lambda)}
      {2\bar p^{\,+}\sqrt{1-\xi^2}}\; 
\widetilde{H}^g(\bar x,\xi;t)
+\frac{\ovl u(p^{\,\prime},\lambda')\Delta^+\gamma_5 u(p,\lambda)}
      {4m\,\bar p^{\,+}\sqrt{1-\xi^2}}\; 
\widetilde{E}^g(x,\xi;t)
\eqpt
\end{eqnarray} 
In the forward limit one has 
\begin{equation} 
\label{eq:pol-forwardgluon}
\widetilde{H}^g(\bar x,\xi=0;t=0)=\bar x\,\Delta g(\bar x)
\eqcm
\end{equation}
with the ordinary polarised gluon distribution $\Delta g(x)$.
Repeating all steps of the derivation presented in
Sect.~\ref{sec:gluonSPD} we obtain the overlap representation of the
polarised gluon SPDs. {}From the combinations of polarisation vectors
in Eqs.~(\ref{eq:polarisation-matrix}) and
(\ref{eq:polarisation-matrix-ERBL}) the antisymmetric tensor
$\varepsilon_{\perp\,\alpha'\alpha}$ now picks up only terms
proportional to $\mbox{sign}(\mu_j)$. Thus, the overlap representation
of polarised gluon SPDs in the region $\xi<\bar x<1$ is given by
(\ref{eq:gluonSPD}) and in the region $-\xi<\bar x<\xi$ by
(\ref{eq:gluonSPD-ERBL}), both completed by a factor
$\mbox{sign}(\mu_j)$ as in (\ref{eq:pol-quarkSPD}) and
(\ref{eq:pol-antiquarkSPD}), and by an additional overall change of
sign, which originates from contracting (\ref{eq:polarisation-matrix})
and (\ref{eq:polarisation-matrix-ERBL}) with
$i\varepsilon^\adj_{\perp\,\alpha'\alpha}$. The SPDs $\widetilde{H}^g$
and $\widetilde{E}^g$ are odd functions in $\bar x$, from which their
overlap representation in the region $-1<\bar x<-\xi$ is readily
deduced.

%%%%%%%%%%%%%%%%%%%%%%%%%%%%%%%%%%%%%%%%%%%%%%%%%%%%
\subsection{Parton helicity changing distributions}

Apart from the unpolarised and polarised skewed distributions
discussed so far, there are also twist-two skewed distributions that
change the helicity of the active parton \cite{Hoodbhoy:1998vm}. The
corresponding quark distributions are constructed from the operator
$\sum_c \bar\psi{}_q^{\,c}\,\sigma^{+i}\gamma_5\,\psi_q^{\,c}$, and
one of them becomes the ordinary quark transversity distribution
$\delta q(x)$ in the forward limit. For the gluons there are skewed
distributions involving a helicity transfer by two units, going with
the tensor $t_{\perp}^{\alpha'\alpha}$ in
Eq.~(\ref{eq:polarisation-matrix}). They appear in deeply virtual
Compton scattering at the $\alpha_s$ level
\cite{Hoodbhoy:1998vm,Diehl:1997bu}.

Our methods can be applied in a straightforward manner to obtain
overlap representations for these SPDs for the entire interval
$-1<\bar{x}<1$. In the regions $\xi<\bar{x}<1$ and $-1<\bar{x}<-\xi$
one no longer has to sum over $\beta=\beta'$ and $j$, as in
Eq.~(\ref{eq:quarkSPD}) and its counterparts, but double sums over
$\beta$, $\beta'$ and $j$, $j'$.  For a given state $\beta$ one must
choose $\beta'$ so as to obtain matching of the quantum numbers for
all spectators, and for the active parton after its helicity has been
changed. Furthermore, a combinatorial factor
$1/\sqrt{\rule{0pt}{1.4ex} n_j n_{j'}}$ appears, where 
$n_j$ ($n_{j'}$) is the number of partons in the LCWF of the initial (final)
proton that have the same discrete quantum numbers as the active
parton $j$ ($j'$).

Notice also that the two active gluons of the skewed gluon distribution 
in the central region $0 < \bar{x} < \xi$ now have the same
helicity and colour. As a consequence, the combinatorial factor 
$1/\sqrt{n_j n_{j'}}$ in the overlap formula is to be replaced with 
$1/\sqrt{n_j (n_j-1)}$, where $n_j$ is the number of gluons in the 
incoming LCWF with the same helicity and colour as the active gluon $j$.

%%%%%%%%%%%%%%%%%%%%%%%%%%%%%%%%%%%%%%%%%%%%%%%%%%%%%%%%%%%%%%%%%%%
\section{General properties of SPDs}
\label{sec:properties}
%%%%%%%%%%%%%%%%%%%%%%%%%%%%%%%%%%%%%%%%%%%%%%%%%%%%%%%%%%%%%%%%%%%
Let us now discuss general properties of the overlap representations
for the matrix elements ${\cal H}_{\lambda\lambda'}$ and
$\widetilde{\cal H}_{\lambda\lambda'}$. Through
Eqs.~(\ref{eq:quark-helcomb}) and (\ref{eq:pol-quark-helcomb}) they
provide linear combinations of the SPDs $H$, $E$ and $\widetilde{H}$,
$\widetilde{E}$, respectively. Evaluating ${\cal H}_{\lambda\lambda'}$
for both proton helicity flip and non-flip, one then obtains $H$, $E$,
$\widetilde{H}$, $\widetilde{E}$ separately for each quark flavour $q$
and for gluons.  At this point we can make a remark on the proton
helicity flip combinations ${\cal H}_{\lambda-\lambda}$ and
$\widetilde{\cal H}_{\lambda-\lambda}$.  The overlap condition
$\beta=\beta'$ in the regions $\xi<\bar{x}<1$ and $-1<\bar{x}<-\xi$
implies that all parton helicities in the initial and final state have
to match, so that they cannot add up to the overall proton helicity in
at least one of the wave functions $\Psi^{-\lambda}$ or
$\Psi^\lambda$.  The same holds in the the region $-\xi<\bar{x}<\xi$
as one can see from (\ref{eq:quarkSPD-ERBL}) and
(\ref{eq:gluonSPD-ERBL}).  It follows that in at least one of them the
orbital angular momentum carried by the partons contributes to the
proton helicity. That $E^q$ and $E^g$ involve parton orbital angular
momentum in an essential way is also reflected in Ji's angular
momentum sum rule \cite{Ji:1998pc}.

In the forward limit the overlaps are solely given by the
contributions from $N\to N$ transitions which now describe the full
interval $-1<\bar x<1$.  As is evident from Eqs.~(\ref{eq:tilde-args})
and (\ref{eq:hat-args}) the arguments of both wave functions are now
identical, and their respective overlaps (\ref{eq:quarkSPD}) and
(\ref{eq:pol-quarkSPD}) for unpolarised and polarised quarks reduce to
the representations of the ordinary parton distribution functions in
terms of LCWFs \cite{Brodsky:1989pv}. With the help of
(\ref{eq:quark-helcomb}) and (\ref{eq:pol-quark-helcomb}) we see that
our overlap representations respect the reduction formulas
\cite{Ji:1997ek,Radyushkin:1997ki}
\begin{eqnarray}
q(\bar x)&=&\sum_N q^{(N)}(\bar x)= H^q(\bar x,0;0) 
\eqcm \nn\\
\Delta q(\bar x)&=&\sum_N \Delta q^{(N)}(\bar x)= 
\widetilde{H}^q(\bar x,0;0)
\eqpt 
\label{eq:red}
\end{eqnarray} 
Likewise, the overlap representations (\ref{eq:antiquarkSPD}) and
(\ref{eq:pol-antiquarkSPD}) reduce in the forward limit to the
representations of the ordinary, unpolarised and polarised, antiquark
distributions. Analogous reduction formulas hold for gluons, see
(\ref{eq:forwardgluon}) and (\ref{eq:pol-forwardgluon}).

The SPDs are related to form factors by sum rules like
\cite{Ji:1997ek} 
\begin{equation}
 F_1^q(t) = \int_{-1}^1 \d\bar{x}\, H^q(\bar{x},\xi;t) \eqcm \qquad
 F_2^q(t) = \int_{-1}^1 \d\bar{x}\, E^q(\bar{x},\xi;t) \eqpt
\label{ff-red}
\end{equation}
Due to Lorentz invariance these relations hold in any reference frame,
and are thus independent of $\xi$. To evaluate the wave function
overlap it is convenient to choose a frame where $\xi=0$ (we will
henceforth call such frames ``symmetric''). Taking the first moment of
our overlap representations, we find the usual Drell-Yan formulae for
form factors~\cite{Drell:1970km}. Symmetric frames are special because
the central region disappears and the LCWF arguments have a purely
transverse shift while $\tilde x_i=\hat x'_i=\bar x_i$, see
(\ref{eq:tilde-args}) and (\ref{eq:hat-args}). In a frame with $\xi=0$
the contribution of quarks of flavour $q$ to the Dirac form factor is
thus related to ${\cal H}^q_{++} = H^q$ (see (\ref{eq:quark-helcomb}))
by
\begin{eqnarray}
F_1^q(t)&=&
\sum_N F_1^{q(N)}(t)=\int_{-1}^1\d\bar x\; H^q(\bar x, 0;t)
\nn\\
&=& 
\sum_{N,\beta} \sum_{j}\;
\delta_{s_j q}\;
\int [\d\bar x]_N [\d^2 \bar{\bf k}_\perp]_N \;
\Psi_{N,\beta}^{*\,+}(\hat r')\,
\Psi_{N,\beta}^+(\tilde r) \eqpt
\label{over0}
\end{eqnarray} 
The Pauli form factor is related to ${\cal H}^q_{-+}$ and thus to
$E^q$ by
\begin{eqnarray}
F_2^q(t)&=&
\sum_N F_2^{q(N)}(t)=\int_{-1}^1\d\bar x\; E^q(\bar x, 0;t)
\nn\\
&=&
\frac{2m}{\eta\sqrt{-t}}\;
\sum_{N,\beta} \sum_{j}\;
\delta_{s_j q}\;
\int [\d\bar x]_N [\d^2 \bar{\bf k}_\perp]_N \;
\Psi_{N,\beta}^{*\,-}(\hat r')\, \Psi_{N,\beta}^+(\tilde r)
\eqpt
\end{eqnarray} 
An expression analogous to (\ref{over0}) relates the axial vector form
factor to $\widetilde{H}^q$ at $\xi=0$. Notice that the pseudoscalar
form factor cannot be represented in the same way, because the
corresponding SPD $\widetilde E^q$ is multiplied in
(\ref{eq:pol-quark-helcomb}) with a factor that vanishes for $\xi=0$.

Wide angle Compton scattering and electroproduction of mesons
can also be described in symmetric frames, and it can be 
shown~\cite{Diehl:1999kh,Radyushkin:1998rt,Huang:2000kd}
that the soft physics in these processes is encoded in new form factors
representing the $1/\bar{x}$-moments of $\xi=0$ SPDs. Examples are
\begin{eqnarray}
R_V^{\,q}(t)&=& \sum_N R_V^{\,q(N)}(t)=\int_{-1}^1
\frac{\d\bar x}{\bar x}\;H^q(\bar x, 0;t)\eqcm \nn\\
R_A^{\,q}(t)&=& \sum_N R_A^{\,q(N)}(t)=\int_{-1}^1 
\frac{\d\bar x}{\bar x}\;
\mbox{sign}(\bar x)\,
\widetilde{H}^q(\bar x, 0;t)
\eqpt
\end{eqnarray} 
Given the relative minus sign between the overlap representations
(\ref{eq:quarkSPD}) and (\ref{eq:antiquarkSPD}) for $H^q$ in the
regions $\xi<\bar x<1$ and $-1<\bar x<-\xi$, active quarks and
antiquarks contribute with opposite signs in $F^q_1(t)$ but with the
same sign in $R^{\,q}_V(t)$. This reflects the different charge
conjugation properties of these form factors. Because there is no
relative minus sign between the corresponding representations
(\ref{eq:pol-quarkSPD}) and (\ref{eq:pol-antiquarkSPD}) for
$\widetilde{H}^q$, a factor $\mbox{sign}(\bar x)$ explicitly appears
in the expression for the Compton form factor $R^{\,q}_A(t)$, where
active quarks and antiquarks contribute again with the same sign.

Except in frames where $\xi=0$, the sum rules (\ref{ff-red}) for form
factors can obviously be satisfied only if the contributions from $\xi
<\bar x <1$, $-1<\bar{x}<-\xi$, and from the central region $-\xi<\bar
x <\xi$, are taken into account.  The neglect of any one leads to
contradictions. That the contribution from the central region may be
substantial, even dominant, can be seen from examples discussed by
Isgur and Llewellyn Smith \cite{isg89}. These authors evaluated
overlap contributions only from the region $\xi <\bar x <1$ in
infinite momentum frames that are obtained from the Breit frame by
either a transverse or a longitudinal boost.  Only the first choice
corresponds to a symmetric frame, while in the second frame
$\Delta_\perp$ is zero and $\xi$ is given by $\xi\simeq 1+ 2m^2/t$ at
large invariant momentum transfer. Thus, not surprisingly after our
discussion, Isgur and Llewellyn Smith arrived at apparently frame
dependent results for the form factors. The missing contributions from
the central region resolve this discrepancy. An explanation of this
discrepancy, which is in line with ours, has already been given by
Sawicki within a covariant approach \cite{Sawicki:1991sr}.

The $\xi$-independence of the sum rules (\ref{ff-red}) is a remarkable
consequence of Lorentz invariance. More generally, higher moments of
SPDs are polynomials in $\xi$ \cite{Ji:1998pc}, 
\begin{equation}
\int_{-1}^1 \d\bar{x} \;\bar{x}^{n-1}\; H^q (\bar{x},\xi;t) 
                 = \sum_{i=0}^{[n/2]} h_{ni}^q(t)\; \xi^{2i} 
\label{poly}
\end{equation}
for positive integer $n$, where $[n/2]$ is the largest integer smaller
or equal to $n/2$. For the overlap representation of the SPDs to
satisfy the polynomiality conditions (\ref{poly}) the LCWFs of higher
Fock states must be related to those of lower Fock states in such a
way that the $\xi$ dependence of the respective contributions from the
$N\to N$ and $N+1\to N-1$ transitions combine in a suitable way. Such
relations are provided by the equations of motion.  Explicit examples
constructed from field theoretical models exhibit this feature
\cite{Sawicki:1991sr,Brodsky:2000}, and it is beyond the scope of this
work to investigate how this works in detail.

Another important property of SPDs is their behaviour at the
transition points, $\bar{x}=\xi$ and $\bar{x}=-\xi$, between the
different regimes in $\bar x$. If they are not continuous at these
points, one will find logarithmically divergent results when
convoluting them with hard-scattering kernels, for instance in deeply
virtual Compton scattering. Whether their first derivatives can be
continuous as well is poorly understood at present, but some models
such as the meson pole contributions discussed in
Sect.~\ref{sec:applications} do lead to discontinuous first
derivatives.  The overlap representation of SPDs might give answers to
these questions, but this will necessitate again an analysis of the
dynamical relations between wave functions for different Fock states
in a hadron.

Taking the limit $\bar x\to \xi$ from above, the momentum fraction
$\hat{x}_j'$ of the quark in the final proton LCWF tends to zero
according to Eq~(\ref{eq:hat-args}). Similarly we see from
Eq.~(\ref{eq:hat-args-central}) that the momentum fraction
$\tilde{x}_{j'}$ of the active antiquark tends to zero if one takes
the limit $\bar x\to \xi$ from below. For $\bar x\to -\xi$ the
situation is analogous. The behaviour of SPDs at these points is thus
related to LCWFs in the limit when one parton momentum fraction goes
to zero. Recent work suggests that LCWFs need not necessarily vanish
in this limit \cite{Antonuccio:1997tw}. Let us however add that even
if they did vanish, one could not conclude that SPDs are zero at $\bar
x=\pm \xi$, since it is not clear that taking the limit $\bar x\to \pm
\xi$ in the SPDs can be interchanged with taking the infinite sum over
the parton number $N$ in Eqs.~(\ref{eq:Focksummation}) and
(\ref{eq:Focksummation-ERBL}). Remember that the limit $x\to 0$ in the
case of ordinary parton distributions 
involves an infinite number of Fock states. The parton distributions
are divergent at this point, whereas each individual Fock state gives
a finite if not vanishing contribution, assuming that LCWFs themselves
do not diverge in the limit where a parton carries zero momentum.

The symmetry of the SPDs under $\xi\to -\xi$ \cite{Ji:1998pc} is
manifest in the overlap representation. As is evident from Eqs.\
(\ref{eq:tilde-args}) and (\ref{eq:hat-args}), the LCWF arguments
$\tilde{r}$ and $\hat{r}'$ are interchanged under the joint
transformation $\xi \to -\xi$ and $\Delta_\perp \to -\Delta_\perp$.
Together with (\ref{eq:quarkSPD}) this leads to
\begin{equation}
{\cal H}^{q(N\to N)}_{\lambda'\lambda} (\bar{x}, \xi, \Delta_\perp)
\,=\, \Big[{\cal H}^{q(N\to N)}_{\lambda\lambda'} 
(\bar{x}, -\xi, -\Delta_\perp)\Big]^*
\label{eq:symm}
\end{equation}
and to similar relations for gluons and for the matrix elements
$\widetilde{{\cal H}}_{\lambda'\lambda}$. As a consequence of
time-reversal invariance the SPDs are real valued, and due to Lorentz
covariance they only depend on $\Delta_\perp$ through its
square. Thus, the insertion of (\ref{eq:symm}) into
(\ref{eq:quark-helcomb}) leads to the required symmetry
\begin{equation}
H^{q(N\to N)}(\bar{x},\xi,t)=H^{q(N\to N)}(\bar{x},-\xi,t)
\end{equation}
in the region $\xi<\bar{x}<1$. Analogous results are obtained for the
other SPDs and for the region $-1<\bar x<-\xi$. In the central region
$-\xi<\bar{x}<\xi$ the transformation $\xi\to -\xi$ turns the
non-diagonal overlaps $N+1\to N-1$ into the appropriate overlaps
$N-1\to N+1$.  Note that the prefactor $\sqrt{1-\xi}^{\,2-N}
\sqrt{1+\xi}^{\,-N}$ in the overlap representation
(\ref{eq:quarkSPD-ERBL}) and in its counterparts for gluons and
polarised SPDs can be written as $\sqrt{1-\xi}^{\,1-(N-1)}
\sqrt{1+\xi}^{\,1-(N+1)}$ so as to exhibit their invariance under the
simultaneous exchange of $\xi\leftrightarrow-\xi$ and
$N+1\leftrightarrow N-1$.

We also remark that the overlap representations of the matrix
elements ${\cal H}_{\lambda\lambda}$ for quarks and gluons satisfy the
positivity constraint in the region $\xi < \bar{x} <1$ derived in
\cite{Pire:1999nw,Radyushkin:1999es}. As we remarked
in \cite{Diehl:1999kh}, Eq.\ (\ref{eq:quarkSPD}) has the structure of
a scalar product in the Hilbert space of LCWFs. Summing over all Fock
states and using Schwarz's inequality and the reduction formulas
(\ref{eq:red}), one immediately finds a bound for the quark matrix
element ${\cal H}_{\lambda\lambda}$,
\begin{equation}
\left| H^{q}(\bar{x},\xi;t) 
       - \frac{\xi^2}{1-\xi^2}\, E^{q}(\bar{x},\xi;t) \right|
     \leq \frac{1}{\sqrt{1-\xi^2}}\;
          \sqrt{q(x_1)\;q(x_2)} \eqcm
\label{eq:qbound}
\end{equation}
with
\begin{equation}
x_1=\frac{\bar x+\xi}{1+\xi}
\eqcm
\qquad
x_2=\frac{\bar x-\xi}{1-\xi}
\eqcm
\end{equation} 
and from (\ref{eq:gluonSPD})
\begin{equation}
\left| H^{g}(\bar{x},\xi;t) 
       - \frac{\xi^2}{1-\xi^2}\, E^{g}(\bar{x},\xi;t)  \right|
     \leq \sqrt{\frac{\bar{x}^2-\xi^2}{1-\xi^2}}\;
          \sqrt{g(x_1)\;g(x_2)} 
\label{eq:gbound}
\end{equation}
for gluons.  Without summing over $N$ one obtains individual bounds
for the contributions from the $N$-particle Fock states to the
respective distributions. The relations (\ref{eq:qbound}) and 
(\ref{eq:gbound}) are precisely the bounds derived in
\cite{Pire:1999nw,Radyushkin:1999es,Diehl:1999kh}, except that the
contributions of $E^q$ and $E^g$ to the proton helicity non-flip
transition have previously been overlooked.  With this proviso, the
weaker bound on $H^g$ obtained by Martin and
Ryskin~\cite{Martin:1998wy} follows directly from
(\ref{eq:gbound}). Using the scalar product structure, one can also
write down the Schwarz inequality for the proton helicity flip matrix
elements ${\cal H}_{\lambda-\lambda}^{q,g}$ and obtains a new bound
\begin{equation}
\frac{\sqrt{t_0-t}}{2m}\;
\Big| E^{q}(\bar{x},\xi;t)  \Big|
     \leq \frac{1}{\sqrt{1-\xi^2}}\;
          \sqrt{q(x_1)\;q(x_2)} 
\eqcm
\label{eq:qbound-new}
\end{equation}
on $E^q$ and a similar bound on $E^{g}$. Of course there are
corresponding bounds for the region $-1<\bar x<-\xi$ as well. 
Finally, one can write down analogous bounds involving the
combinations $H\pm \widetilde{H}$ and $E\pm \widetilde{E}$ for quarks
or gluons with a definite helicity, along with the corresponding
combinations $q \pm \Delta q$ or $g \pm \Delta g$ of the ordinary
parton distributions.

We finally note that, as the usual parton distributions, SPDs depend 
on a factorisation scale $Q^2$ in a way that can be calculated in 
perturbation theory and is expressed in evolution 
equations~\cite{Muller:1994fv,Ji:1997ek,Radyushkin:1997ki}. In 
the overlap representation, this dependence shows up in the
fact that both the LCWFs themselves and the integrals over 
the ${\bf k}_\perp$ of the partons in their overlap have to be 
regulated in the ultraviolet, with $Q^2$ playing the role of 
characeristic momentum scale in the regulator~\cite{Brodsky:1989pv}. How 
such a regularisation can be implemented in detail and how it leads to 
the well-known evolution equations for SPDs is beyond the scope of our study.

%%%%%%%%%%%%%%%%%%%%%%%%%%%%%%%%%%%%%%%%%%%%%%%%%%%%%%%%%%%%%%%%%%%
\section{Phenomenological applications}
\label{sec:applications}
%%%%%%%%%%%%%%%%%%%%%%%%%%%%%%%%%%%%%%%%%%%%%%%%%%%%%%%%%%%%%%%%%%%

Most of the existing phenomenological applications of the overlap
representations are carried through in symmetric frames and concern
the electromagnetic form factors at large momentum transfer, for the
proton, e.g.\
\cite{Diehl:1999kh,Radyushkin:1999es,isg89,Bolz:1996sw,Carlson:1987sw},
and for the pion, e.g.\
\cite{isg89,Jakob:1993iw,Kisslinger:1993cm}.\footnote{ In LCWF based
constituent quark models one also uses overlap formulas like
(\ref{over0}) (see e.g.~\cite{Cardarelli:1995dc}) restricted, for
obvious reasons, to the valence Fock state contributions.}.  Since
little is known about LCWFs with nonzero orbital angular momentum,
proton helicity flip is mostly ignored, and only results for the Dirac
form factor $F_1(t)$ are obtained.  Attempts have also been made to
model SPDs through the overlap of LCWFs, including the limiting case
of the ordinary parton distribution functions \cite{Diehl:1999kh}.
Again, the description of proton helicity flip is more difficult and
has so far been shunned. We remark that often there is some
justification to neglect the contributions of the SPDs $E$ and
$\widetilde{E}$ to scattering processes. From (\ref{eq:quark-helcomb})
and (\ref{eq:pol-quark-helcomb}) we see that they appear with
prefactors $\xi^2 /(1-\xi)^2$ or $\sqrt{t_0-t}/(2m)$ that are
typically small in the kinematics considered. We emphasise that at
small $t$ $E$ itself \emph{cannot}\ be small compared to $H$: its first
moment builds up the Pauli form factor, which is by no means small
compared to the Dirac form factor close to $t=0$. Also, the pion pole
contribution to $\widetilde{E}$ can be so large that this distribution
cannot be neglected in processes where it contributes
\cite{Vanderhaeghen:1999xj}.

The overlaps are usually evaluated from soft LCWFs, representing full
wave functions with their perturbative, large $k_\perp$ tails
removed. Typically, the transverse momentum dependence of the soft
LCWFs is parameterised as a Gaussian
\begin{equation}
\Psi^\lambda_{N\beta} \sim \exp{\left[-a_N^2\, 
                   \sum^N_{i=1} k_{\perp i}^2/x_i\right]} \eqpt
\label{gaussian}
\end{equation}
The soft overlaps evaluated from such wave functions are large and
dominate the form factors at momentum transfer of the order of 10
GeV$^2$ while the competing perturbative QCD contributions, which may
be viewed as the overlaps of the perturbative tails of the LCWFs,
provide only minor corrections of perhaps less than $10\%$ in the case
of the proton \cite{Bolz:1996sw,Bolz:1995hb} and about $20 - 40\%$ in
the case of the pion \cite{Jakob:1993iw,Braun:2000uj}. For
asymptotically large $t$ the perturbative contributions will take the
lead, and the soft overlaps merely represent power corrections to
them.\footnote{ For soft LCWFs of the Gaussian type
(\protect\ref{gaussian}) the overlap contributions to the Dirac form
factor fall off as $F_1^{q(N)} \propto t^{-2(N+l_g-1)}$ where $l_g$ is
the number of gluons in the Fock state.}

With a few exceptions, e.g.~\cite{Diehl:1999kh,Bolz:1996sw}, overlap
contributions have so far only been evaluated for the valence Fock
states.  This is expected to be a reasonable approximation for form
factors at large $t$ and for parton distributions at large $\bar
x$. Higher Fock states can be taken into account for instance by
assuming that (\ref{gaussian}) holds for all Fock states with a common
transverse size parameter $a=a_N$. This simplification allows one to
sum over $N$ explicitly and, without need for specifying the
$x$-dependences of the LCWFs, to relate the results to the usual
parton distributions. One thus obtains a very simple model for form
factors and the underlying SPDs at $\xi=0$,
\cite{Diehl:1999kh,Radyushkin:1999es,Vogt:2000ku,Barone:1993ej} which 
nicely demonstrates the link between exclusive
and inclusive hard scattering reactions. For quarks of flavour $q$,
for instance, the $\xi=0$ proton SPDs read
\begin{eqnarray}
H^q(\bar{x}, 0;t)&=&\, \exp{\left[\,\frac12\, a^2 t\,
               \frac{1-\bar{x}}{\bar{x}}\right]}\,\; q(\bar{x})
\eqcm 
\nn\\
\widetilde{H}^q(\bar{x}, 0;t)&=&\, \exp{\left[\,\frac12\, a^2 t\,
               \frac{1-\bar{x}}{\bar{x}}\right]}\, \Delta q(\bar{x})
\eqcm
\label{xi0}
\end{eqnarray}
within that model. Analogous expressions hold for gluons
\cite{Huang:2000kd}. Taking the parton distributions from one of the
current analyses of deep inelastic lepton-nucleon scattering, e.g.\
\cite{Gluck:1998xa}, and using a value of 1 GeV$^{-1}$ for the
transverse size parameter, $a$, one can evaluate the form factors
$F_1$, $R_V$ and $R_A$ from the SPDs (\ref{xi0}). Fair agreement with
experiment is obtained
\cite{Diehl:1999kh,Radyushkin:1999es,Diehl:1999tr}. In
\cite{Vogt:2000ku} an SPD like (\ref{xi0}) is considered as a model
that is valid at a low scale. The use of DGLAP evolution equations
allows then to evaluate this SPD at larger scales and to explore the
scale dependence of the transverse size parameter. The relation of
$\xi=0$ SPDs to the spatial distribution of partons inside hadrons is
discussed in \cite{Burkardt:2000za}.

It is to be stressed that the overlap representations of the SPDs, we
have given are exact. In other words, provided all Fock state wave
functions are known, the leading-twist SPDs can be constructed from
their overlaps. Although the Fock state decomposition in principle
already comprises meson pole contributions, which become manifest in
the overlap for $-\xi<\bar{x}<\xi$, it is not easy to built them up in
a phenomenological ansatz of the LCWFs. Thus, with regard to the
prominent role of the pole terms in some processes and in particular
kinematical regions, it might be of advantage in phenomenology to add
them explicitly. An example is set by the pion pole in the case of the
proton SPDs. Since the pion couples to protons as $\bar{u} \gamma_5 u$
it contributes only to the SPD $\tilde{E}^q$.  With regard to the
pion's isovector nature the $\pi^0$-pole contribution at small $t$,
near the pole reads
\cite{Radyushkin:1999es}
\begin{equation}
\tilde{E}^u_{pole}(x,\xi;t) = - \tilde{E}^d_{pole}(x,\xi;t) =
       \theta(-\xi \leq \bar{x} \leq \xi)\, \frac{g_\pi}{m_\pi^2-t}\,
       \frac{1}{\xi} \;
       \varphi_\pi(\tau)
\eqcm
\label{eq:pipole}
\end{equation}
where $m_\pi$ and $g_\pi$ are the pion mass and a coupling constant,
respectively. $\varphi_\pi$ is the pion distribution 
amplitude, i.e., its valence Fock state wave function 
integrated over transverse
momentum. Its argument is the momentum fraction the quark
carries with respect to the pion momentum and is related to the
variables of the SPD, $\bar{x}$ and $\xi$, by 
\begin{equation}
\tau= \frac{\bar{x}+\xi}{2 \xi}
\eqpt
\end{equation}
The calculation of the SPDs at $\xi\neq 0$ requires contributions from
the central region (see e.g.\ Eq.\ (\ref{eq:quarkSPD-ERBL})). With the
exception of the flavour non-diagonal $b-u$ SPDs appearing in
$B\to\pi$ transitions \cite{Feldmann:2000sm} the full $\xi\neq 0$ SPDs
have not been calculated from the overlap representation as yet, owing
to insufficient phenomenological experience with the required higher
Fock state wave functions. In our previous work \cite{Diehl:1999kh} we
therefore presented only results for the SPDs in the regions $\xi
<\bar{x} <1$ and $-1<\bar{x}<-\xi$, evaluated from LCWF of the
Gaussian type (\ref{gaussian}). The situation for the $b-u$ SPDs is
special, because one expects the $B$-meson LCWFs to be strongly peaked
for momentum fractions of the $b$-quark given by the ratio of
$b$-quark and $B$-meson mass.  As a consequence, all overlap
contributions with the exception of the valence contribution from the
region $\xi <\bar{x} <1$ are suppressed by inverse powers of the
$b$-quark mass. Therefore the valence contribution to the region
$\xi<\bar x<1$, together with the $B^*$-resonance contribution to the
central region, parameterised in analogy to (\ref{eq:pipole}), makes
up most of the $b-u$ SPDs. The SPD approach thus allows a
superposition of resonance and overlap contributions without a
matching procedure.

%%%%%%%%%%%%%%%%%%%%%%%%%%%%%%%%%%%%%%%%%%%%%%%%%%%%%%%%%%%%%%%%%%%
\section{Summary}
\label{sec:summary}
%%%%%%%%%%%%%%%%%%%%%%%%%%%%%%%%%%%%%%%%%%%%%%%%%%%%%%%%%%%%%%%%%%%

LCWFs provide a convenient way to describe the quark and gluon
structure of hadrons in QCD. They naturally appear in the Fock state
decomposition of hadron states within the context of LC quantisation,
and they are the non-perturbative input that describes hadron
structure in many hard exclusive processes.  In a different class of
exclusive reactions, the nonperturbative physics is contained in more
complex quantities, namely in skewed parton distributions.  In the
present paper we have used the Fock state decomposition to derive the
representation of SPDs through the overlap of LCWFs, which can be seen
as the more elementary quantities. Our method can be applied to both
quark and gluon distributions, including their various spin
combinations.

The overlap representation readily allows us to interpret SPDs in the
physics picture of the parton model. Usual parton distributions, being
constructed from squared hadron wave functions, represent classical
probabilities to find a specified parton within a hadron. In contrast,
skewed distributions are interference terms between the wave functions
for different parton configurations---this is one way to see why they
contain more information on the hadron's structure than the usual
distributions alone. Our derivation naturally leads to the result that
in the central region $-\xi<\bar x<\xi$ of the SPDs the overlap is
between wave functions for Fock states with different parton
number. The overlap representation also makes it transparent that the
proton spin-flip distributions $E$ and $\tilde{E}$ intrinsically
involve the orbital angular momentum carried by the partons.

The overlap formulae directly reflect general properties of the SPDs,
in particular their connection to the usual parton distributions and
to hadronic form factors. Also, their crossing symmetry under $\xi\to
-\xi$ is manifest. Writing SPDs as a overlap also provides an elegant
way to derive their positivity bounds in the regions $\xi<\bar{x}<1$
and $-1<\bar{x}<\xi$.  Other features of SPDs are less immediate in
this representation: the polynomiality property (\ref{poly}) and the
behaviour at the points $\bar x=\pm \xi$ involve the nontrivial
relationship between the wave functions for different Fock states that
is due to the equation of motion, an issue we did not investigate in
the present work.

Finally, the overlap representation provides us with strategies to
model SPDs and their moments in kinematical regions where only a
limited number of hadron Fock states is important. Examples studied in
the literature concern form factors at large momentum transfer in most
cases, e.g.\ the electromagnetic ones of the pion and the proton,
transition form factors and those specific to wide angle Compton
scattering and electroproduction of mesons. First attempts to evaluate
SPDs in the region $\xi<\bar x<1$ and at $\xi=0$ from the overlap
representation can also be found.

In this work we have been concerned with twist-two SPDs, which have a
straightforward interpretation in the framework of LC
quantisation. The SPDs at twist-three level have recently been
classified and investigated \cite{Anikin:2000em}. They appear for
instance in the $1/Q$ power corrections to deeply virtual Compton
scattering. A representation of such distributions in terms of LCWFs
should also be possible, as has been achieved for the usual
twist-three spin-dependent structure functions in
\cite{Mankiewicz:1991az}.

%%%%%%%%%%%%%%%%%%%%%%%%%%%%%%%%%%%%%%%%%%%%%%%%%%%%%%%%%%%%%%%%%%%
\section*{Acknowledgments}
%%%%%%%%%%%%%%%%%%%%%%%%%%%%%%%%%%%%%%%%%%%%%%%%%%%%%%%%%%%%%%%%%%%

We wish to express our gratitude to Paul Hoyer for his continued
encouragement of this work and for his careful reading of the manuscript.

%%%%%%%%%%%%%%%%%%%%%%%%%%%%%%%%%%%%%%%%%%%%%%%%%%%%%%%%%%%%%%%%%%%
\begin{appendix}
\section*{Appendix}
%%%%%%%%%%%%%%%%%%%%%%%%%%%%%%%%%%%%%%%%%%%%%%%%%%%%%%%%%%%%%%%%%%%

In this Appendix we discuss an alternative choice of kinematical
variables, where the three-momentum $\bf{p}$ of the initial proton is
along the ${\bf e}_3$-axis (see Fig.~\ref{fig:one}(b)). We present our main
results in terms of this alternative set of kinematical variables. The
momenta $p$ and $p'$ characterising the initial and final hadron state
can be parameterised as
\begin{eqnarray}
p  &=& \lcvec{p^+}
             {\frac{m^2}{2p^+}}
             {{\bf 0}_\perp} \eqcm \nn\\ 
p^{\,\prime} &=&  \lcvec{(1-\zeta)\,p^+}
              {\frac{m^2+{\bf\Delta}_\perp^2}{2(1-\zeta)p^+}}
              {{\bf\Delta}_\perp}
\label{eq:A-rad-parameterisation}
\end{eqnarray}
with the skewedness parameter 
\begin{equation} 
\zeta=\frac{(p-p')^+}{p^+} 
\eqpt
\end{equation}
The difference of hadron momenta takes the form
\begin{equation} 
\Delta=p^{\,\prime}-p=\lcvec{-\zeta\,p^+}
                  {\frac{\zeta m^2+{\bf\Delta}_\perp^2}{2(1-\zeta)p^+}}
                  {{\bf\Delta}_\perp} 
\eqcm
\end{equation} 
and with the parametrisation (\ref{eq:A-rad-parameterisation}) its
square reads $t=-(\zeta^2m^2+{\bf\Delta}_\perp^2)/(1-\zeta)$. Positivity 
of ${\bf\Delta}_\perp^2$
implies a minimal value $-t_0=(\zeta^2 m^2)/(1-\zeta)$ at given
$\zeta$, or correspondingly, a maximum allowed value for $\zeta$ at
given $t$.

Let us first consider the quark SPDs defined in terms of the
alternative set of kinematical variables by
\begin{eqnarray} 
\lefteqn{F_{\lambda'\lambda}^{\,q} \equiv 
\frac{1}{2\sqrt{1-\zeta}} \;
\sum_c
\int\frac{\d z^-}{2\pi}\;e^{i\,x p^+z^-}\;
\langle p^{\,\prime},\lambda'|
        \,\bar\psi_q^{\,c}(0)\,\gamma^+\,\psi_q^{\,c}(\bar z)\,
                                        |p,\lambda\rangle}
\nn\\[1\baselineskip]
&=&
 \frac{\overline u(p^{\,\prime},\lambda')\gamma^+ u(p,\lambda)}
      {2\,p^+\,\sqrt{1-\zeta}}\; 
\widetilde{\cal F}^{\,q}_\zeta(x;t)
+\frac{\overline u(p^{\,\prime},\lambda')
         {i\sigma^{+\alpha}\Delta_\alpha} u(p,\lambda)}
      {4\,m\,p^+\,\sqrt{1-\zeta}}\; 
\widetilde{\cal K}^{\,q}_\zeta(x;t)
\end{eqnarray} 
for unpolarised quarks, and
\begin{eqnarray} 
\lefteqn{G_{\lambda'\lambda}^{\,q} \equiv
\frac{1}{2\sqrt{1-\zeta}} \;
\sum_c
\int\frac{\d z^-}{2\pi}\;e^{i\,x p^+z^-}\;
\langle p^{\,\prime},\lambda'|
   \,\bar\psi_q^{\,c}(0)\,\gamma^+\gamma_5\,\psi_q^{\,c}(\bar z)\,
                                                |p,\lambda\rangle} 
\nn\\[1\baselineskip]
&=&
 \frac{\overline 
       u(p^{\,\prime},\lambda')\gamma^+\gamma_5 u(p,\lambda)}
      {2\,p^+\,\sqrt{1-\zeta}}\;
\widetilde{\cal G}^{\,q}_\zeta(x;t)
+\frac{\overline 
       u(p^{\,\prime},\lambda')\Delta^+\gamma_5 u(p,\lambda)}
      {4\,m\,p^+\,\sqrt{1-\zeta}}\;
\widetilde{\cal P}^{\,q}_\zeta(x;t)
\eqcm
\end{eqnarray} 
which defines the polarised SPDs. The SPD $\widetilde{\cal
F}^{\,q}_\zeta(x;t)$ is related to the previously defined quark SPD by
(see~\cite{Radyushkin:1997ki,Ji:1998pc})
\begin{equation} 
(1+\xi)\,H^q(\bar x)=
\widetilde{\cal F}_{\zeta(\xi)}^{\,q} \left(x(\bar x,\xi)\right)
\end{equation} 
with
\begin{equation} 
\label{eq:A-xvonxji}
x(\bar x,\xi)=\frac{\bar x+\xi}{1+\xi} 
\;,\qquad
\zeta(\xi)=\frac{2\xi}{1+\xi}
\eqpt
\end{equation}
Analogous relations hold for $\widetilde{\cal K}^{\,q}_\zeta(x;t)$,
$\widetilde{\cal G}^{\,q}_\zeta(x;t)$ and 
$\widetilde{\cal P}^{\,q}_\zeta(x;t)$. The definitions for the three 
different
kinematical regions, reexpressed in terms of the alternative
variables, are $\zeta<x<1$ and $-1+\zeta<x<0$ for the regions with
$N\to N$ transitions, and $0<x<\zeta$ for the central region. Let us
first consider quark SPDs in the region $\zeta<x<1$. A frame where the
parametrisation (\ref{eq:A-rad-parameterisation}) for the hadron
momenta holds is already a hadron-in frame. Thus, the arguments $r_i$
of the LCWFs for the incoming hadron are given by the plus components
and transverse parts of the parton momenta. To determine the arguments
of the LCWF for the outgoing hadron one applies the transverse boost
(\ref{eq:plustrafo}) with the parameters $b^+=(1-\zeta)p^+$ 
and $\bf{b}_\perp=\bf{\Delta}_\perp$ leading to a hadron-out frame. We
label quantities in the hadron-out frame with a breve. From the
spectator condition
\begin{equation} 
k'_i=k^\adj_i \eqcm
\qquad \mbox{for }i\neq j 
\end{equation} 
together with momentum conservation, one obtains relations between the
arguments of the LCWF for the outgoing hadron (with breve) to the ones
for the incoming hadron:
\begin{eqnarray} 
\breve x'_i= \frac{x_i}{1-\zeta} \eqcm &\qquad& 
\breve{\bf k}'_{\perp i}={\bf k}^{\phantom{.}}_{\perp i}
                        -\frac{x_i}{1-\zeta}\,
                         {\bf\Delta}^{\phantom{.}}_\perp
\;,\qquad \qquad \mbox{for }i\neq j \eqcm \nn\\
\breve x'_j=\frac{x_j-\zeta}{1-\zeta} \eqcm &\qquad& 
\breve{\bf k}'_{\perp j}={\bf k}^{\phantom{.}}_{\perp j}
                        +\frac{1-x_j}{1-\zeta}\,
                         {\bf\Delta}^{\phantom{.}}_\perp
                        \eqpt
\label{eq:A-breve-args}
\end{eqnarray} 
The overlap representation for the unpolarised quark SPDs in the
region $\zeta<x<1$ takes the form
\begin{eqnarray} 
F_{\lambda'\lambda,\,\zeta}^{q(N\to N)}(x;t)&=&
\sqrt{1-\zeta}^{\,1-N} \;
\sum_{\beta=\beta'}\;
\sum_{j} \delta_{s_j q}\;
\nn\\
&\times&
\int [\d x]_N [\d^2 {\bf k}_\perp]_N \;
\delta\left(x-x_j\right)\;
\Psi_{N,\beta'}^{*\,\lambda'}(\breve r')\,
\Psi_{N,\beta}^\lambda(r)
\eqcm
\label{eq:A-quarkSPD}
\end{eqnarray} 
and an analogous equation with the additional factor
$\mbox{sign}(\mu_j)$ holds for the matrix element
$G_{\lambda'\lambda}^{q(N\to N)}$ involving the polarised quark SPDs
. Summation over N leads to the full quark SPDs in the region
$\zeta<x<1$.

The overlap representation for the quark SPDs in the region
$-1+\zeta<x<0$, describing the emission and reabsorption of an
antiquark, is obtained from (\ref{eq:A-quarkSPD}) or from its analogue
for the polarised SPDs by the replacement of the $\delta$-function by
$\delta(x-\zeta+x_j)$, and by a change of sign for the unpolarised
antiquark SPD. The arguments of the LCWFs are again to be taken as
specified in (\ref{eq:A-breve-args}).

Finally we consider the region $0<x<\zeta$, where the SPDs describe a
hadron emitting a quark-antiquark pair. By combination of the
spectator condition with momentum conservation one obtains relations
between the arguments of the LCWF for the outgoing hadron to the ones
of the LCWF for the incoming hadron of the form
(\ref{eq:A-breve-args}) for $i\neq j,j'$, and additional relations
between the relative momentum coordinates of active quark and
antiquark
\begin{equation}
x_{j'}=-(x_j-\zeta) \eqcm \qquad
{\bf k}_{\perp j'}=-{\bf k}_{\perp j'}-{\bf\Delta}^{\phantom{.}}_\perp
\eqpt
\end{equation} 
The overlap representation for unpolarised quark SPDs in the region
$0<x<\zeta$ for the $N-1\to N+1$ transition now reads
\begin{eqnarray}
F_{\lambda'\lambda,\,\zeta}^{q(N+1\to N-1)}(x;t)&=&
\sqrt{1-\zeta}^{\,2-N}\;
\sum_{\beta,\beta'}\;
\sum_{j,j'=1}^{N+1}
\frac{1}{\sqrt{n_j n_{j'} \rule{0pt}{0.7em}}}\;
\delta_{{\bar s}_{j'}s^\adj_j}\;\delta_{s^\adj_j q}\;
\delta_{\mu_{j'}-\mu^\adj_j}\,
\delta_{c_{j'}c^\adj_j}\;
\nn\\
&\times&
\prod_{i=1\atop i\ne j,j'}^{N+1} 
\delta_{s'_i s^\adj_i}\;
\delta_{\mu_i'\mu^\adj_i}\;
\delta_{c'_i c^\adj_i}\;
\int \d x_j
\prod_{i=1\atop i\ne j,j'}^{N+1} 
\d x_i\;
\delta\left(\sum_{i=1\atop i\ne j,j'}^{N+1}x_i-1+\zeta\right)\;
\nn\\
&\times&
\int \d^2 {\bf k}_{\perp j}
\prod_{i=1\atop i\ne j,j'}^{N+1} \d^2 {\bf k}_{\perp i} \;
(16\pi^3)^{1-N}\;
\delta^{(2)}\!\left(\sum_{i=1\atop i\ne j,j'}^{N+1}
            {\bf k}_{\perp i}-{\bf\Delta}_\perp\right)
\nn\\[\baselineskip]
&\times&
\delta\left(x-x_j\right)\;
\Psi_{N-1,\beta'}^{*\,\lambda'}(\breve r')\,
\Psi_{N+1,\beta}^\lambda(r)\;
\eqcm
\label{eq:A-quarkSPD-ERBL}
\end{eqnarray}
and similar for $G_{\lambda'\lambda,\,\zeta}^{q(N+1\to N-1)}(x;t)$ with
the additional factor $\mbox{sign}(\mu_j)$.

Now we turn to the gluon SPDs which are defined by
\begin{eqnarray}
F^g_{\lambda'\lambda} &\equiv& 
\frac{-g_{\perp,\alpha'\alpha}}{p^+\sqrt{1-\zeta}}\;
\sum_c
\int \frac{\d z^-}{2\pi} e^{ix\,p^+z^-} \;
\langle p',\lambda'|
        \,G_c^{+\alpha'}(0)\,G_c^{+\alpha}(\bar z)\,
                                  |p,\lambda\rangle
\nn\\[1\baselineskip]
&=&
 \frac{\ovl u(p',\lambda')\gamma^+ u(p,\lambda)}
      {2\,p^+\,\sqrt{1-\zeta}}\;
\widetilde{\cal F}^g_\zeta(x;t)
+\frac{\ovl u(p',\lambda')
       i\sigma^{+\alpha}\Delta_\alpha u(p,\lambda)}
      {4\,m\,p^+\,\sqrt{1-\zeta}}\;
\widetilde{\cal K}^g_\zeta(x;t)
\end{eqnarray}
for the unpolarised case, and
\begin{eqnarray} 
G^g_{\lambda'\lambda} &\equiv& 
\frac{i\varepsilon^{~}_{\perp\, \alpha'\alpha}}
     {p^+\sqrt{1-\zeta}}\;
\sum_c
\int \frac{\d z^-}{2\pi} e^{ix\,p^+z^-} \;
\langle p',\lambda'|\,
      G_c^{+\mu}(0)\,G_c^{+\nu}(\bar z)\,|p,\lambda\rangle
\nn\\
&=&
 \frac{\overline 
       u(p^{\,\prime},\lambda')\gamma^+\gamma_5 u(p,\lambda)}
      {2\,p^+\,\sqrt{1-\zeta}}\; 
\widetilde{\cal G}^g_\zeta(x;t)
+\frac{\overline 
       u(p^{\,\prime},\lambda')\Delta^+\gamma_5 u(p,\lambda)}
      {4\,m\,p^+\,\sqrt{1-\zeta}}\; 
\widetilde{\cal P}^g_\zeta(x;t)
\end{eqnarray}
for the polarised gluon SPDs. The SPD $\widetilde{\cal
F}^{\,g}_\zeta(x;t)$ is related to the previously defined gluon SPD by
\begin{equation} 
(1+\xi)\,H^g(\bar x)=
\widetilde{\cal F}_{\zeta(\xi)}^{\,g} \left(x(\bar x,\xi)\right)
\end{equation} 
with the transformations (\ref{eq:A-xvonxji}). Analogous relations
hold for $\widetilde{\cal K}^{\,g}_\zeta(x;t)$, 
$\widetilde{\cal G}^{\,g}_\zeta(x;t)$ and 
$\widetilde{\cal P}^{\,g}_\zeta(x;t)$.

The overlap representations for the gluon SPDs in the region
$\zeta<x<1$ can be readily obtained from the equation for quark SPDs
(\ref{eq:A-quarkSPD}) by adding a factor $\sqrt{x(x-\zeta)}$ for the
unpolarised, or $-\mbox{sign}(\mu_j)\sqrt{x(x-\zeta)}$ for the
polarised SPD. Likewise, one obtains the gluon SPDs for the 
$N+1\to N-1$ transitions in the region $0<x<\zeta$ from
Eq.~(\ref{eq:A-quarkSPD-ERBL}) by adding a factor $\sqrt{x(\zeta-x)}$
in the unpolarised, or $\mbox{sign}(\mu_j)\sqrt{x(\zeta-x)}$ in the
polarised case. Note that the unpolarised (polarised) gluon SPDs are
even (odd) under the interchange $x\to \zeta-x$.

\end{appendix}
%%%%%%%%%%%%%%%%%%%%%%%%%%%%%%%%%%%%%%%%%%%%%%%%%%%%%%%%%%%%%%%%%%%

%%%%%%%%%%%%%%%%%%%%%%%%%%%%%%%%%%%%%%%%%%%%%%%%%%%%%%%%%%%%%%%%%%%

%%%%%%%%%%%%%%%%%%%%%%%%%%%%%%%%%%%%%%%%%%%%%%%%%%%%%%%%%%%%%%%%%%%

\newpage

\begin{center}
{\bf \Large Erratum}\\[2\baselineskip]

{\it The Overlap Representation of Skewed Quark and Gluon
Distributions}\\[\baselineskip ]

Nucl.\ Phys.\  {\bf B596} (2001) 33-65\\[0.5\baselineskip]

M.~Diehl, Th.~Feldmann, R.~Jakob and P.~Kroll\\[2\baselineskip]
\end{center} 

\begin{itemize}
\item The polarised skewed gluon distributions as defined in Eqs.~(69)
and (100) have the wrong sign: in the forward limit they tend to
$-\bar x\,\Delta g(\bar x)$, instead of $\bar x\,\Delta g(\bar x)$
as stated in Eq.~(70). We recall that the forward polarised gluon
density $\Delta g(\bar x)$ gives the difference of right-handed and
left-handed gluons in a right-handed proton, and remark that its
definition in Eq.~(4.46) of the CTEQ `handbook of pQCD' (Reference
[17] in the text) has the wrong sign, too~\cite{collins}.
      
To have the correct forward limit $\bar x\,\Delta g(\bar x)$ of the
polarised skewed gluon distributions, the order of the indices in the
tensor $\varepsilon^{~}_{\perp\, \alpha'\alpha}$ has to be changed
into $\varepsilon^{~}_{\perp\, \alpha\alpha'}$ in Eqs.~(69) and (100).
Accordingly, in the paragraph after Eq.~(70) the remark about an
additional overall change of sign should be dropped. For the same
reason, in a statement at the very end of the Appendix two signs
should be exchanged and it should read correctly
\begin{quote} 
``\ldots The overlap representations for the gluon SPDs in the region
$\zeta<x<1$ can be readily obtained from the equation for quark SPDs
(96) by adding a factor $\sqrt{x(x-\zeta)}$ for the unpolarised, or
$+\mbox{sign}(\mu_j)\sqrt{x(x-\zeta)}$ for the polarised
SPD. Likewise, one obtains the gluon SPDs for the $N+1\to N-1$
transitions in the region $0<x<\zeta$ from Eq.~(98) by adding a factor
$\sqrt{x(\zeta-x)}$ in the unpolarised, or
$-\mbox{sign}(\mu_j)\sqrt{x(\zeta-x)}$ in the polarised case. Note
that \ldots''
\end{quote} 

\item In Eq.~(28) a factor $\sqrt{2}$ is missing on the right hand
side.
\end{itemize}

\end{document}